\documentclass[
11pt,
a4paper,
letterpaper,
abstracton,
]{scrartcl}

\usepackage[ngerman, english]{babel}
\usepackage[latin1]{inputenc}
\usepackage[T1]{fontenc}
\usepackage{graphicx}
\usepackage{float}
\usepackage[section]{placeins}
\usepackage{subcaption}
\usepackage{url}
\usepackage{marvosym}
\usepackage{lmodern}
\usepackage{newclude}
\usepackage{xcolor}
\usepackage[doublespacing]{setspace}
\usepackage{ragged2e} 
\usepackage{enumitem}	
\usepackage{tabularx}	
\usepackage{longtable}
\usepackage{booktabs}	
\usepackage{dcolumn}	
\usepackage{threeparttable}
\usepackage{pdflscape} 	
\usepackage{amsmath, amssymb}
\usepackage[official]{eurosym} 
\usepackage{caption}          
\usepackage[left=3cm, right=3cm, bottom=3cm, top=3cm]{geometry} 
\usepackage{graphicx} 
\usepackage{mathtools} 
\usepackage{mathpazo} 
\usepackage{array}  
\usepackage[title]{appendix}

\newcolumntype{C}[1]{>{\centering\let\newline\\\arraybackslash\hspace{0pt}}m{#1}}
\newcolumntype{L}[1]{>{\raggedright\let\newline\\\arraybackslash\hspace{0pt}}m{#1}}

\usepackage[pdftex,bookmarks=true, colorlinks=false, pdfborder={0 0 0}]{hyperref} 	

\usepackage{fancyref}

\newcommand*{\fancyrefapplabelprefix}{app}
\fancyrefaddcaptions{english}{%
	\providecommand*{\frefappname}{appendix}%
	\providecommand*{\Frefappname}{Appendix}%
}
\frefformat{plain}{\fancyrefapplabelprefix}{\frefappname\fancyrefdefaultspacing#1}
\Frefformat{plain}{\fancyrefapplabelprefix}{\Frefappname\fancyrefdefaultspacing#1}

\usepackage[natbibapa]{apacite}
\usepackage{doi}

\newenvironment{literatur}{%
	\parskip0pt \parindent0pt \justifying 
	\def\lititem{\hangindent=1cm \hangafter1}}{%
	\par\ignorespaces}

\begin{document}
	
	\begin{singlespace}
		
		\title{Internal migration after a uniform minimum wage introduction \vspace{0.1cm}}
		\author{Alexander Moog \\
			\small{\emph {JGU Mainz}} 
			\thanks{Johannes Gutenberg-University Mainz (JGU Mainz), Address: JGU Mainz, Jakob-Welder-Weg 4, 55128 Mainz, e-mail: amoog@uni-mainz.de \\
				I thankfully acknowledge Thorsten Schank for helpful comments and fruitful discussions.}
		}
		
		\date{April 2024}
		
		\maketitle
		\thispagestyle{empty}
		
		\begin{abstract}
			\noindent Internal migration is an essential aspect to study labor mobility. I exploit the German statutory minimum wage introduction in 2015 to estimate its push and pull effects on internal migration using a 2\% sample of administrative data. In a conditional fixed effects Poisson difference-in-differences framework with a continuous treatment, I find that the minimum wage introduction leads to an increase in the out-migration of low-skilled workers with migrant background by 25\% with an increasing tendency over time from districts where a high share of workers are subject to the minimum wage (high-bite districts). In contrast the migration decision of native-born low-skilled workers is not affected by the policy. However, both native-born low-skilled workers and those with a migrant background do relocate across establishments, leaving high-bite districts as their workplace. In addition, I find an increase for unemployed individuals with a migrant background in out-migrating from high-bite districts. These results emphasize the importance of considering the effects on geographical labor mobility when implementing and analyzing policies that affect the determinants of internal migration. \emph{[172 words]}   
			
			\vspace{0.3cm}
			\noindent \textbf{JEL Classification}: J08, J61, J64, R23  \\ 
			\noindent \textbf{Keywords}: Minimum wages, internal migration, geographic labor mobility,  immigrant workers, unemployment 
		\end{abstract}
	\end{singlespace}
	
	\clearpage
	\setcounter{page}{1}
	
	\section{Introduction}\label{sec:introduction}
	
	Internal migration is an essential part in the study of geographic labor mobility and the resulting spatial socio-demographic distribution of the labor force. In Germany two internal migration patterns dominated shifts in the socio-demographic distribution, namely the east-west migration after the reunification in 1990 (see, e.g., \citet{rosenbaum2022}) and migration across the urban-rural sphere (see, e.g., \citet{stawarz2019}). Both internal migration patterns altered the socio-demographic structure, particularly in former East German rural regions, with older individuals remaining and younger individuals leaving \citep{rosenbaum2022}. \citet{stawarz2019} also note that urbanization and subsequent sub-urbanization into economically more prosperous regions accelerated the aforementioned socio-demographic shift in rural East German regions, as well as in rural West German regions. Both phases of internal migration are a result of changes in socio-economic spatial differences. Thereby, internal migration is described as dominantly driven by economic factors such as relatively higher (expected and short-term) regional unemployment, leading individuals to migrate out of a region -- push factor -- and relatively higher (expected) regional income, leading individuals either to move into a region or to stay in a region -- pull or restraining factor (see, e.g., \citet{schuendeln2009, bauer2019, mitze2012}). Given regional unemployment and wages to be the driving forces of internal migration poses the question to which extent internal migration patterns shift if a policy is introduced which possibly affects both regional unemployment and wages, namely the German statutory minimum wage introduction in 2015.
	
	Minimum wages are a popular policy tool aimed at reducing poverty and economic inequality by increasing wages at the lower end of the wage distribution. In addition, according to neo-classical labor market theory, minimum wages could have a negative effect on employment, leading to rising unemployment in regions where minimum wages are introduced or increased.\footnote{In monopsonistic and oligopsonistic labor markets, imposing minimum wages theoretically can lead to positive employment effects (see, e.g., \citet{bhaskar2002}).} \citet{monras2019} developed a spatial model to analyze the impact of minimum wage increases on the internal migration of low-wage workers. The study identifies two main theoretical mechanisms for migration, namely increasing regional wages as a pull factor and increasing unemployment as a push factor for low-wage workers. In the United States, he finds an increase in the out-migration of low-skilled workers\footnote{\citet{monras2019} employs low-skilled workers as a proxy group to represent workers who are subject to the minimum wage.} in states where minimum wages are introduced or increased, which is mainly driven by increases in regional unemployment due to minimum wage increases, i.e. the push factor dominates the pull factor. However, the United States are a patchwork of different minimum wage laws at the federal, state, and county level, making it difficult to compare to Germany, where a uniform minimum wage of 8.50 Euro was introduced in 2015. While a uniform minimum wage equalizes regional wages of all workers who were earning below the minimum wage before its introduction, the share of workers subject to the minimum wage can vary across regions. In Germany, approximately 10\% to 15\% of workers were subject to the minimum wage introduction. In rural districts\footnote{Districts refers to the NUTS-3 regional classification. Germany is organized in 400 districts.} of former East Germany, up to 30\% of workers were affected by the minimum wage, compared to those working in former West Germany who were only slightly affected.\footnote{For reference see panel (A.1) in \fref{fig:soc_econ} in \fref{sec:t_labdem}.}\footnote{For the remainder of this paper, I will refer to regions with a high share of workers subject to the minimum wage as high-bite regions and, correspondingly, to regions with a low share of workers subject to the minimum wage as low-bite regions. The term "bite" refers to how deeply the introduction of the minimum wage has bitten into the regional wage distribution.} Given the strong regional variation at the district level in workers subject to the minimum wage, I adapt \citet{monras2019} theory to a uniform minimum wage introduction. I show that under a uniform minimum wage, the migration decision of workers subject to the minimum wage is  determined by regional differences in changes in unemployment rates, where relatively higher increases in regional unemployment rates induce affected workers to move to regions with relatively lower increases in unemployment rates. Thus, in regions where relatively more workers are subject to the minimum wage by also experiencing a relatively high increase in regional unemployment rate, out-migration should increase more than in regions where fewer workers are subject to the minimum wage. In addition to theory, \citet{dustmann2021} found an increase in the closure of small establishments in German high-bite municipalities\footnote{As of 31.12.2021 Germany is organized in 10,994 municipalities. Municipalities are a subset of the regional classification NUTS-3, i.e. districts.} and a subsequent increase in labor mobility from small establishments to large and more "stable" establishments, which are characterized by having more than 250 employees and a low turnover rate. Therefore, the theoretical effect on out-migration could be driven by (short-term) unemployment due to the closure of small establishments, which is not visible in aggregated annual data, and the subsequent reallocation to corresponding large establishments located in low-bite regions. Moreover, I show that unemployed individuals may begin to migrate from high-bite regions when the reservation wage is lower than the minimum wage. Also, I motivate by evidence from \citet{cadena2014} for the United States that the effect may be stronger for individuals with migrant background than for native-born, since he found that individuals with migrant background are more responsive to changes in labor market conditions.
	
	To estimate the effect, I aggregate the weakly anonymized version of the Employment Panel of Integrated Employment Histories (SIAB) (Graf et al., 2023) at the district-year level. This allows me to measure the number of in- and out-migrants by district per year. Given the nature of the respective variables as count variables, I employ a difference-in-differences specification with continuous treatment, which is commonly used in the minimum wage literature (see, e.g., \citet{card1992}), within a conditional fixed effects Poisson model. I separately estimate the effect of the uniform minimum wage introduction in Germany on the number of inflows to and outflows from districts.
	
	This paper contributes to the existing literature in several dimensions. Firstly, it provides further evidence for the theoretical framework proposed by \citet{monras2019}, including its application in the context of a uniform minimum wage introduction. To provide empirical evidence, I apply \citet{wooldridge2021, wooldridge2022} recent theoretical work on non-linear difference-in-differences, which supports the application of a difference-in-differences framework in a conditional fixed effects Poisson model. By separately estimating the treatment effects of the minimum wage introduction on the number of inflows and outflows for workers subject to the minimum wage, it is possible to distinguish between the push and pull effects of the policy intervention. Thus, this approach provides an appropriate way to causally study internal migration responses to policy interventions when count data is available.
	
	Secondly, I provide novel and detailed evidence on the effect of the introduction of the German minimum wage in 2015 on internal migration. I find no effect on inflows, but I show an increase in average out-migration of low-skilled workers with a migrant background and a corresponding reallocation between establishments across districts. In contrast, I find no internal migration effect for natives, but an increase in reallocation between establishments across districts. Thus, this study confirms the findings of \citet{dustmann2021} regarding the increase in labor mobility across establishments and commuting out of high-bite regions, but only for native-born workers.  Therefore, I provide evidence for Germany which supports \citet{cadena2014} findings for the United States, that suggest that individuals with a migrant background respond more strongly to changes in local labor market conditions by moving longer distances than natives. Furthermore, I discuss that the differences in migration decisions between natives and immigrants may be due to different reasons. Native-born individuals may have stronger personal ties to their home district compared to those with a migrant background, which could explain why the latter group may react more strongly to changing economic conditions. Additionally, network effects among individuals with similar migrant backgrounds could contribute to the increasing tendency for outflows over time following the introduction of a minimum wage.
	
	Thirdly, I provide theoretical considerations and novel empirical results on the internal migration decision of unemployed individuals in case of a uniform minimum wage introduction. I find an increase in out-migration of unemployed individuals with a migrant background due to the minimum wage introduction, but with a two-years lag. However, the reasons why unemployed individuals choose to migrate are not entirely clear. According to theoretical considerations, unemployed individuals whose reservation wage falls below the minimum wage after its introduction may follow the same migration patterns as workers subject to the minimum wage. Also, unemployed individuals may also follow the migration decision of their partners who are subject to the minimum wage and out-migrate.
	
	The remainder is structured as follows. \Fref{sec:t} theoretically motivates the effect of a uniform minimum wage introduction on internal migration and formulates the main research hypotheses. \Fref{sec:int_mig} presents the main dataset, describes the main identification strategy and analyzes and discusses respective results. \Fref{sec:conclusion} concludes. 
	
	\section{Internal migration and a uniform minimum wage introduction}\label{sec:t}
	
	I start this section, by motivating possible economic channels of the minimum wage introduction on internal migration by theory and former evidence in \fref{sec:t_labdem} to \fref{sec:t_migbac}. I conclude this chapter by providing a summary of key objectives and hypotheses relevant for the analysis in \fref{sec:t_rq}.
	
	\subsection{Regional labor demand and spatial reallocation of workers}\label{sec:t_labdem}
	
	To illustrate theoretical effects of the uniform minimum wage introduction in Germany in 2015 on internal migration, I adapt the theory by \citet{monras2019} of a corresponding effect of non-uniform minimum wage introductions and increases on the geographical mobility of low-wage workers. An individual's decision to move to another region is rationally driven by the expected utility they perceive. If the expected utility in another region exceeds the current expected utility the individual would rationally decide to move. \citet{monras2019} models a worker's indirect utility function for each region $s$ as a function between receiving unemployment benefits or going to work as follows\footnote{Assuming constant price levels across regions and a reservation wage of 0.}. 
	\begin{align}\label{eq:equat1}
		V_{s} = u_{s}B^{\rho}+(1-u_{s})(1-\tau)^{\rho}w_s^{\rho}
	\end{align}
	where $u_s$ is the regional probability of being unemployed, $B$ are uniform unemployment benefits, $\tau$ the tax rate, $\rho$ a measure of risk aversion and $w_s$ is the regional wage. Given $B < (1-\tau)w_s$, the decision to move is fully captured by spatial differences in $u_s$ and $w_s$.\footnote{Unemployment benefits are assumed to be lower than net wages. This makes sense since workers would not work if unemployment benefits exceeded their respective net wages.} Thus, the lower the unemployment rate in a region or the higher the regional wage, the higher the indirect utility of living in this region. In contrast to \citet{monras2019}, I analyze a uniform minimum wage introduction, which simplifies the model to the extent that after the introduction, spatial wages of workers affected by the minimum wage ($w^{min} > w_s$) increase and equalize across regions, such that $w_s = w^{min}$. This allows me to analyze the indifference condition directly. For simplicity, I illustrate potential effects on internal migration in a two-regions world where the indifference condition is $V_1 = V_2$.\footnote{This can easily be extended to $n$ regions.} An individual considers moving from region 1 to region 2 either if regional wages in region 2 increase stronger than in region 1 or if the unemployment rate in region 1 increases stronger than in region 2 such that region 2 becomes relatively more attractive than region 1. To illustrate the effect of the minimum wage introduction, I assume region 1 to be a high-bite region and region 2 to be a low-bite region such that $w_1 < w_2 < w_{min}$. When holding unemployment rates constant, high-bite regions become relatively more attractive, since $\Delta w_1 > \Delta w_2$ due to the minimum wage introduction. Therefore, inflows into high-bite regions increase. However, a minimum wage introduction can also increase the unemployment rate due to negative employment effects. Hence, an individual living in region 1 would move to region 2 if $\Delta u_{1}$ is sufficiently higher than $\Delta u_{2}$ such that the gain in utility due to increasing wages is reversed due to the increase in the unemployment rate. Therefore, the effect on internal migration could be driven by negative employment effects in high-bite regions induced by the introduction of the minimum wage. Thus, the more workers subject to the minimum wage in a region, the greater the potential increase in a region's unemployment rate due to firms reducing or substituting labor. If a region's unemployment rate increases, regions with relatively lower increases in unemployment rates become more attractive. As a result, workers will move out of regions that are highly affected by the minimum wage. In addition, if a region's labor demand is relatively more elastic, the effect may be stronger.\footnote{Labor demand elasticities may differ across regions depending on the composition of firms in a region. Labor demand may be more elastic in regions with relatively more multinational or larger firms, as these firms are more likely to substitute labor by capital or outsource labor (see e.g. \citet{popp2023}) In contrast, smaller firms are less likely to substitute or reduce labor when wages rise and vice versa. Therefore, labor demand is less elastic in regions with a high share of smaller firms.}
	
	However, significant negative aggregate employment effects of the minimum wage introduction have not been reported in the literature, but a recent strand of the German minimum wage literature shows reallocation effects of workers across establishments \citep{dustmann2021, haelbig2023}.\footnote{\citet{bossler2024} found a negative employment effect for the subgroup of individuals working in marginal employment due to the minimum wage introduction. However, I exclude marginal employed workers from my main analysis.} For example, while not reporting average employment effects, \citet{dustmann2021} find reallocation effects of low-wage workers between small and "better" establishments due to the introduction of the minimum wage. Here, "better" establishments are characterized as large and stable, i.e. with more than 250 employees and a low turnover rate. The reported effect is mainly due to the closure of small establishments located in high-bite municipalities. However, the closure of small establishments in high-bite regions may have led to very short-term unemployment -- within one year -- which is not observable in annual data, but may be perceived by workers as an increase in (short-term) unemployment. On an annual basis, this effect may only have been observable as a reallocation effect between establishments. This leads to a similar reasoning with respect to the out-migration from high-bite regions, since smaller establishments close and (very) short-term or perceived unemployment increases relatively more in high-bite regions than in low-bite regions. Moreover, the decision to move is now also based on the location of "better" establishments. Thus, "better" establishments may be located in more distant regions, making commuting unfeasible and thus inducing the worker to move.\footnote{Conversely, \citet{dustmann2021} found an increase in commuting to other municipalities. However, the authors did not estimate a potential regional reallocation effect of affected workers.} Subsequently, migration out of high-bite regions may increase due to the closure of small establishments and the location of "better" establishments in regions where commuting from the origin is not feasible.
	
	\begin{figure}[!tbh]
		\caption{Minimum wage bite and concentration of big establishments in 2014}\label{fig:soc_econ}
		\centering
		\includegraphics[width=1\textwidth]{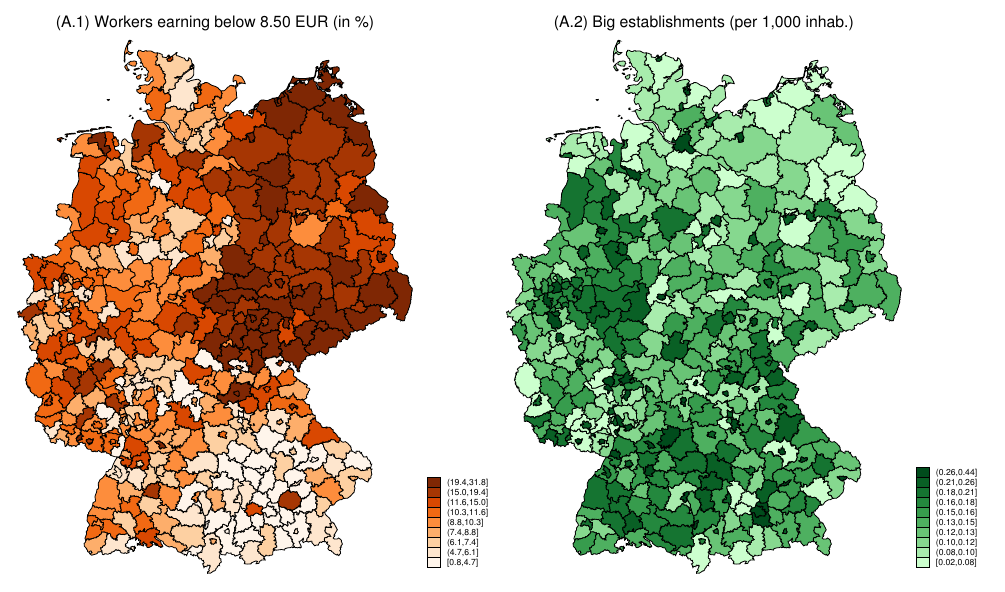}
		\justifying
		\begin{singlespace}
			\noindent \footnotesize \textit{Note}: Workers earning below 8.50 EUR for establishments with more than 10 employees: mean=10.84\%, sd=6.01\%; Big establishments (per 1,000 inhabitants): mean=0.159, sd=0.071. Big establishments are defined as establishments with more than 250 employees. \textit{Source: Structure of earnings survey (SES) and Regionaldatenbank (2024a), own calculations}  
		\end{singlespace}
	\end{figure}
	
	\Fref{fig:soc_econ} presents the share of workers subject to the minimum wage (A.1) and the regional concentration of establishments with more than 250 employees per capita (A.2), both for the year 2014. Comparing (A.1) and (A.2) indicates a negative spatial correlation between the concentration of large establishments and high-bite regions. This is consistent with reallocation effects as described by \citet{dustmann2021} from high-bite districts into low-bite districts since big establishments seem to be located mainly in low-bite districts. However, big establishments are located more densely in former West German districts whereas high-bite districts are mainly located in former East Germany. Therefore, this indicates that the reallocation effect might not only increase commuting flows from high-bite districts to low-bite districts, but also increases internal migration out of high-bite districts into low-bite districts, since distances seem not feasible for commuting. This further supports the main argumentation for an increase in out-migration from high-bite districts by likewise increasing in-migration into low-bite districts due to the introduction of the minimum wage in Germany in 2015.
	
	\subsection{Spatial reallocation of unemployed}\label{sec:t_unemp}
	By relaxing the assumption $B < (1-\tau)w_s$ and allowing a reservation wage greater than zero, one can use the indirect utility function of \fref{eq:equat1} to describe possible migration patterns of unemployed individuals due to the introduction of a uniform minimum wage. In this simple framework, unemployed individuals whose reservation wage ($w^{reserv}$) becomes less than $w_s$ would relocate to regions where wages exceed the current wage and the reservation wage in order to start working. In the case of a uniform minimum wage, two types of unemployed can be distinguished. First, there is a group of unemployed where $w^{reserv} > w^{min}$ still holds after the minimum wage is introduced, who will not start working anyway. Second, there is a group of unemployed where $w^{reserv} > w_s$ changes to $w^{reserv} < w^{min}$ after the minimum wage is introduced. Since no lower bound spatial wage differentials remain after the minimum wage is introduced, these unemployed would follow a similar migration pattern as low-wage workers, determined by regional unemployment rates and respective labor demand elasticities, as described in \fref{sec:t_labdem}. This group is newly entering the labor market and did not move prior to the introduction of the minimum wage due to wage differentials. Thus, out-migration from high-bite regions of individuals who were unemployed before the minimum wage was introduced may increase after the minimum wage introduction similar as for low-wage workers.
	
	\subsection{Spatial reallocation of individuals with migrant background}\label{sec:t_migbac}
	
	The literature on migration shows that in the United States immigrants choose their destination based on differences in regional labor market conditions, particularly in terms of expected earnings. \citet{cadena2014} reports empirical evidence that individuals with short-term migration backgrounds move longer distances and are more responsive to changes in labor market conditions. He shows that short-term migrants working in low-wage employment tend to be more sensitive to earnings and more flexible in their location choices when the minimum wage increases in a region. Thus, the effect on out-migration described in \fref{sec:t_labdem} and \fref{sec:t_unemp} may be stronger for individuals with a migrant background than for natives, as they are more sensitive to changes in regional labor market conditions affected by the introduction of the minimum wage.
	
	\subsection{Key objectives}\label{sec:t_rq}
	
	In summary, I will examine whether the introduction of a uniform minimum wage in Germany in 2015 increases outflows of workers subject to the minimum wage introduction from high-bite regions. For this purpose, I hypothesize that the outflows from high-bite regions of respective workers increase due to a uniform minimum wage introduction, while the inflow to high-bite regions stays constant or even decreases. Moreover, I will exploit whether parts of the unemployed population do also out-migrate from high-bite regions similar to the response of low-wage workers. Finally, I will examine whether the effects are stronger for individuals with migrant background.  
	
	\section{Data and descriptive statistics}\label{sec:data}
	
	In this section, I first introduce the main data source and important steps in data preparation to reflect internal migration flows. Second, I show relevant descriptive statistics for the internal migration flow dataset.
	
	\subsection{Data source, sample selection and data preparation}\label{sec:dataprep1}
	
	To analyze the effect of the introduction of the minimum wage on migration flows, I use microdata on individual employment histories from the Employment Panel of Integrated Employment Histories (SIAB) of the Institute for Employment Research (IAB) (Graf et al., 2023).\footnote{For more information on the dataset refer to the dataset manual \citep{SIAB_DESC}.} It covers the years from 1975 to 2021 and is a 2\% representative sample of administrative data covering employees subject to social security contributions and the unemployed population in Germany. It excludes civil servants and self-employed. However, it covers the majority of the German labor force.\footnote{It covers about 80\% of the German labor force. For comparison see, e.g. Microcensus 2022.} The dataset merges spells of social security records with spells of unemployment records from the Federal Employment Agency. Most importantly, the SIAB contains information on the district of residence of an individual and, if employed, the district in which the individual works -- geographical information is availahle on NUTS-3 level. In order to construct a yearly panel, I use December 31 as the cut-off date for each year, and in case of parallel spells on the cut-off date, I consider the spell with the highest wage as the main spell for the respective observations\footnote{In general, I follow the data preparation procedure in the manner of \citet{stuber2023guide} to construct a yearly panel.}. I also exclude workers in marginal employment from my main specification because it seems unlikely that such workers would change their district of residence if they are marginally employed in their main job, given the unfeasible costs of moving. However, since marginally employed workers are likely to be strongly affected by the introduction of the minimum wage, I will use them for sensitivity analysis. In addition, I exclude interns, working students, apprentices, family workers, sailors, irregular workers, and (part-time) retirees, as their decision to move is unlikely to be strongly influenced by the valuation of expected wage differentials, as they are likely to value mainly non-pecuniary motives in their occupational choice.    
	
	Based on the micro panel, I construct an aggregate panel with the dimensions district $s$ and year $t$ which includes inflows into ($Inf_{s,t}$) and outflows from ($Out_{s,t}$) a district for a given year. Both of these variables are measured as follows.
	\begin{align}\label{eq:aggregation}
		Inf_{s,t}=\sum_{i}^{N}I\left[s_{i,t} \neq s_{i,t-1}\right]_{i,s,t}~~\text{or}~~Out_{s,t}=\left[\sum_{i}^{N}I\left[s_{i,t} \neq s_{i,t+1}\right]_{i,s,t}\right]_{t-1}
	\end{align}
	To capture inflows, I first construct an indicator variable $I$ that is 1 if an individual changes district $s$ between $t-1$ and $t$ and 0 otherwise. Then I sum the indicator variable over all individuals living in district $s$ and year $t$. Thus, the sum reflects the number of inflows to a district for a year. Outflows are constructed similarly, except that the indicator variable is 1 if an individual changes their district of residence between $t$ and $t+1$, and 0 otherwise. Because of the cut-off date, the sum measures the number of outflows for the year $t+1$ in the period $t$. Therefore, the value calculated for year $t-1$ of the outflow variable must be considered in year $t$. Measurement error could occur if employees report their new residence to their employers with some delay. However, the annual structure should limit this bias. In addition, it is not clear whether individuals report their first or second residence to the employer, which could lead to incorrectly assign individuals to a district. However, this would only pose a problem if individuals switched from their first to their second residence without actually moving, which I expect to be a very rare event. To also analyze whether workers change the district in which they work as a result of the minimum wage, I follow the same procedure for changes in the district of the workplace. This reflects geographic reallocation effects across establishments.
	
	In order to examine heterogeneity according to theory, I measure inflows and outflows based on whether individuals have a migrant background or are native-born. The SIAB contains information on an individuals citizenship where I classify individuals with a German citizenship as native-born and individuals with other citizenships having a migrant background. I further distinguish between being unemployed and being employed in a low-skilled or medium/high-skilled occupation.\footnote{An individual's unemployment status or occupational skill level refers to the period before the move, i.e., the status may change after the move. Therefore, the flows are based on the respective state in period $t-1$.} In addition, I distinguish between flows where the distance between district centroids is greater or less than 150 km, flows within or outside a labor market region, and flows across or within the borders of former East and West Germany in order to analyze potential regional shifts in the regional labor force at different geographic levels. 
	
	\subsection{Descriptive statistics}\label{sec:desc1}
	
	\begin{table}[htb]
		\centering
		\caption{Sample composition before and after the minimum wage introduction }
		\label{tab:tab_socdem}
		\begin{tabular}{lcccc}
			\toprule
			& \multicolumn{2}{c}{Native-born} & \multicolumn{2}{c}{Migrant Background} \\
			\cmidrule(lr){2-3} \cmidrule(lr){4-5}
			& 2010-2014 & 2015-2019 & 2010-2014 & 2015-2019 \\
			\midrule
			Mean age (in years) & 41.482 & 42.508 & 38.993 & 38.397 \\
			Unemployment & 0.115 & 0.090 & 0.274 & 0.260 \\
			Low-skill employment & 0.289 & 0.300 & 0.480 & 0.542 \\
			Imputed daily wage, deflated (2015) & 104.559 & 112.200 & 86.742 & 87.947 \\
			N & 2,760,000 & 2,850,000 & 274,000 & 424,000  \\
			\bottomrule
		\end{tabular}
		\vspace{1mm}
		\begin{singlespace}
			\noindent \justifying \footnotesize \textit{Notes}: This table refers to the full micro data population after excluding marginally employed, interns, working students, apprentices, family workers, sailors, irregular workers and (part-time) retirees. Censored (top-coded) wages are imputed according to \citet{stuber2023guide} who follow a similar procedure as in \citet{dustmann2009} and \citet{card2013}. \textit{Source: SIAB, own calculations}.
		\end{singlespace}
	\end{table}
	
	\Fref{tab:tab_socdem} presents the sample composition in terms of age, unemployment, low-skilled employment and real daily wages of the underlying micro data separately for native-born individuals and those with a migrant background. The table also shows the averages for both groups separately for five years before and five years after the introduction of the minimum wage. Age is rather stable in both samples over time. Unemployment, however, decreases for both samples after the introduction of the minimum wage, following the general decreasing trend of unemployment in Germany for the respective years. Interestingly, for native-born, the share of low-skilled employment increases only slightly, while real daily wages increase on average by 7.3\% in the five years after the introduction of the minimum wage compared to the five years before. In comparison, the average share of low-skilled employees with a migrant background increased by about 6.2 percentage points after the introduction of the minimum wage, while the real daily wage increased only marginally by about 1.4\%. The increase in the number of low-skilled employees with a migrant background could be due to the immigration waves in 2015 and 2016, which can also be observed in the increase in sample size of about 150,000 in the population of individuals with a migrant background. However, this would only be a problem for identification of the treatment effects if migrants were systematically distributed more often to high-bite regions after their arrival. The distribution mechanism of newly arrived immigrants is based on the "K\"onigsteiner Schl\"ussel", which is also used to calculate a state's contribution to common financing in relation to the state's tax revenue and population size. Normally, within states, immigrants are distributed among municipalities relative to their population size. Thus, the distribution of immigrants should not pose a threat to the identification strategy I describe in \fref{sec:ident1}.
	
	\begin{table}[htb]
		\centering
		\caption{Average inflows and outflows by sub-group before and after the minimum wage introduction }
		\label{tab:tab_acc_dat}
		\begin{tabular}{lcccc}
			\toprule
			& \multicolumn{2}{c}{Mean} & \\
			\cmidrule(lr){2-3}
			& {2010-2014} & {2015-2019} & {$\Delta$} \\
			\midrule
			Inflows (native-born) & 41.176 & 38.245 & -2.931 \\
			Inflows (native-born, low-skill employment) & 8.693 & 8.527 & -0.166 \\
			Inflows (native-born, unemployed) & 4.646 & 3.918 & -0.728 \\
			Inflows (migrant background) & 3.955 & 7.590 & 3.635 \\
			Inflows (migrant background, low-skill employment) & 1.204 & 2.982 & 1.778 \\
			Inflows (migrant background, unemployed) & 0.815 & 1.868 & 1.053 \\
			\midrule
			Outflows (native-born) & 41.178 & 38.170 & -3.008 \\
			Outflows (native-born, low-skill employment) & 8.711 & 8.532 & -0.179 \\
			Outflows (native-born, unemployed) & 4.607 & 3.845 & -0.762 \\
			Outflows (migrant background) & 3.954 & 7.665 & 3.711 \\
			Outflows (migrant background, low-skill employment) & 1.185 & 2.978 & 1.793 \\
			Outflows (migrant background, unemployed) & 0.853 & 1.942 & 1.089 \\
			\bottomrule
		\end{tabular}
		\vspace{1mm}
		\begin{singlespace}
			\noindent \justifying \footnotesize \textit{Notes}: Inflows $Inf_{s,t}$ and outflows $Out_{s,t}$ measure the number of individuals (given restrictions) who move into our out of districts per year according to \fref{eq:aggregation}. \textit{Source: SIAB, own calculations}           
		\end{singlespace}
	\end{table}
	
	\Fref{tab:tab_acc_dat} presents the average number of annual inflows and outflows for relevant sub-samples over the five years before the minimum wage was introduced, compared with the corresponding average over the five years after the minimum wage was introduced. Net migration -- inflows minus outflows -- is close to zero for all sub-samples, ensuring that the data generation process is robust to error and not biased by panel attrition in the micro sample.\footnote{Since inflows is a backward-looking measure while outflows is a forward-looking measure (as described in \fref{sec:dataprep1}), slight differences occur due to individuals changing their primary citizenship between 2014 and 2015 from German to any other nationality and vice versa. For example an individual changes their citizenship from French to German between 2014 and 2015 and moves between district $a$ and $b$. Therefore, the move counts as inflow to $b$ for the native-born sub-sample and as outflow from $a$ for the migrant background sub-sample. Changing the citizenship only pose a threat to identification if changing the nationality would count as additional inflow or outflow for districts $b$ and $a$, respectively, which is not the case here.} Average annual outflows decreased for native-born individuals after the minimum wage introduction, while the decrease is close to zero for low-skilled workers. However, the out-migration of individuals with a migrant background increased by about 93.8\% after the introduction of the minimum wage, with the out-migration of low-skilled workers and the unemployed more than doubling. Both increases in outflows could be partly due to the increase in sample size due to the migration waves in 2015 and 2016, and partly due to the introduction of the minimum wage in 2015. Therefore, I will empirically test whether part of the increase in the outflows of individuals with a migrant background is due to the introduction of the minimum wage. For natives, it remains unclear given the negative development of outflows. The decrease in out-migration could be driven by a decrease in out-migration in low-bite districts, while out-migration in high-bite districts increases due to the introduction of the minimum wage, which is not distinguishable in the aggregate data.
	
	\section{Identification strategy}\label{sec:ident1}
	
	In order to estimate potential effects of the minimum wage introduction on internal migration, I make use of a difference-in-differences specification with a continuous treatment that varies at the regional level in the manner of, for instance, \citet{card1992}, which is widely used in the minimum wage literature to study changes in economic outcomes due to the introduction of a minimum wage. The continuous treatment variable $Bite_{s}$ is measured between 0 and 1 as the percentage of employees per district earning less than 8.50 Euros in the last pre-treatment year 2014 per district, i.e., it reflects the treatment intensity by measuring how deep the minimum wage introduction has bitten into the regional wage distribution.\footnote{For the geographical distribution of the minimum wage bite see panel (A.1) in \fref{fig:soc_econ}.} Thus, I exploit the regional variation in the share of workers affected by the minimum wage by comparing changes in migration flows in more affected districts to changes in migration flows in less affected districts to identify the treatment effect on respective outcome variables. For identification the bite must be conditionally exogenous to changes in migration flows and therefore changes in wages in post-treatment periods, since expected wages in post-treatment periods affect the moving decisions of workers earning below the minimum wage threshold in 2014 and also unemployed individuals, as discussed in \fref{sec:t}. There may be also anticipation effects on wages, i.e., treatment effects on wages are measurable for the year 2014 because the minimum wage has already been publicly debated and the law has passed in the respective year. This could pose a threat to conditional exogeneity, since the minimum wage bite would already influence an individual's decision to move in pre-treatment periods. However, the findings of \citet{bossler2023} show an anticipation effect on wages only for the 20th percentile of the wage distribution, though, most likely due to collective bargaining. Given the limited evidence for anticipation effects on wages and since the decision to move is a consequence of the effect on wages, I find it convincing that the conditional exogeneity conditions for the minimum wage bite defined in 2014 hold.
	
	\begin{figure}[ptbh]
		\caption{Distribution of inflows and outflows (2010 - 2019)}\label{fig:hist_move}
		\centering
		\includegraphics[width=1\textwidth]{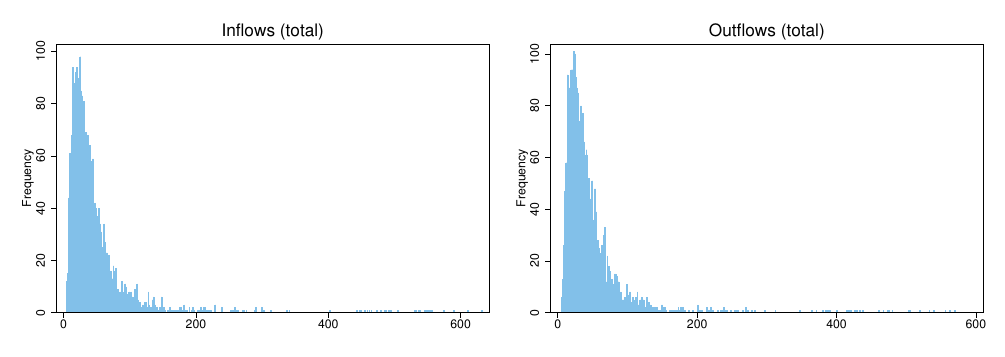}\\
		\begin{singlespace}
			\justifying \noindent \footnotesize \textit{Note}: This figure shows the distribution of the variables $Inf_{s,t}$ and $Out_{s,t}$ for all workers subject to social security and unemployed individuals. For the distribution of respective variables for further sub-groups used in the analysis refer to \fref{app:apx_dist}. Source: SIAB, own calculations.
		\end{singlespace}
	\end{figure}
	
	The main dependent variables are measured as non-negative integers reflecting the number of individuals moving into ($Inf_{s,t}$) or out of ($Out_{s,t}$) a district $s$ within a year $t$ as described in \fref{sec:dataprep1}. As is common for count data, the distribution is centered around small values with a long tail but also includes zeros (see \fref{fig:hist_move}). The distribution is mainly due to districts having different population sizes, with the majority of districts having relatively small populations, while a few district have relatively large populations, both with corresponding numbers of inflows and outflows relative to the population size. Given the nature of the respective dependent variables, I employ an "event-study"-type difference-in-differences specification in a pooled Poisson model of the following form.
	\begin{align}\label{eq:DIDpos}
		P(y_{s,t}=h_{s,t}|\textbf{x}_{s,t}) = \frac{exp[-z(\textbf{x}_{s,t})][z(\textbf{x}_{s,t})]^{h_{s,t}}}{h_{s,t}!},\enspace h_{s,t}=0,1,2,\dots
	\end{align} 
	where $y_{s,t}$ denotes either $Inf_{s,t}$ or $Out_{s,t}$ and $z(\textbf{x}_{s,t})$ is the conditional mean, which includes the difference-in-differences specification as follows.
	\begin{align}
		E(y_{s,t}|\textbf{x}_{s,t}) = exp\left(\alpha+\beta Bite_s + \sum_{t=2010, t\neq2014}^{2019}\delta_{t} Year_{t} + \sum_{t=2010, t\neq2014}^{2019}\gamma_{t} Bite_{s}\cdot Year_{t}\right)  
	\end{align}
	Here, the coefficients of interest are $\gamma_{t}$ -- for both graphical and numerical analysis -- which identify the average treatment effects of the minimum wage introduction on the number of inflows to and outflows from districts as the percentage change given the treatment intensity. The advantage of using a Poisson model is that it produces consistent estimates regardless of the underlying distribution. Compared to estimating a log-linear model, the Poisson model takes into account changes in flows from or to zero. These changes are omitted in the log-linear model because a logarithm is not defined for zero. As is common practice, zero values are typically replaced with a small number before log-transforming, however, which is not recommendable since it leads to inconsistent estimates \citep{silva2006log}. Moreover, \citet{silva2006log} show that log-linear specifications lead to biased estimates under heteroskedasticity even when estimating a fixed effects model. To further underline the choice of the Poisson model, I compare estimates from a log-linear model with fixed effects to the Poisson estimates in \fref{sec:robust1}. The Poisson estimator also allows for over- and underdispersion and serial correlation as long as cluster-robust standard errors are estimated. Crucial to Poisson models, however, is the assumption that the conditional mean is correctly specified. To examine that it is correctly specified, I perform a RESET-type test for each specification as suggested by \citet{wooldridge1999}. To do so, I first estimate the Poisson model with the conditional mean specified as in \fref{eq:DIDpos} and compute the predicted values. I then re-estimate the model by additionally including a quadratic and a cubic term of the predicted values and perform a simple Wald test on both estimators. If they are jointly insignificant, this indicates that the model is correctly specified.  
	
	Due to the presence of panel data and because it is likely that district-level time-invariant factors influence both an individual's decision to move into or out of a district and differences between high- and low-bite areas, I use a conditional fixed effects Poisson model. This additionally controls for respective (short-term) time-invariant unobserved heterogeneity. Time-invariant factors that the model accounts for are, for instance, the infrastructure of a district, where poor infrastructure may cause individuals to move out of a district. Moreover, poor infrastructure is a potential determinant of relatively lower wages compared to regions with a good infrastructure, which would lead to biased estimates in the pooled model. Therefore, I employ a conditional fixed effects Poisson model with the conditional mean specified as follows. 
	\begin{align}\label{eq:DIDposFE}
		\begin{split}
			E(y_{s,t}|c_{s},\textbf{x}_{s,2010},\dots,&\textbf{x}_{s,2019}) = \\\
			&c_{s}exp\left(\sum_{t=2010, t\neq2014}^{2019}\delta_{t} Year_{t} + \sum_{t=2010, t\neq2014}^{2019}\gamma_{t} Bite_{s}\cdot Year_{t}\right)
		\end{split}
	\end{align}
	In contrast to the pooled case, the conditional mean now additionally depends on other time-invariant unobserved factors $c_{s}$ at the district level. All assumptions remain identical to the pooled Poisson estimator and average treatment effects are also estimated by $\gamma_{t}$ given the treatment intensity. Thus, it also allows for any kind of variance-mean relationship -- overdispersion and underdispersion do not lead to biased standard errors -- and it allows for serial correlation as long as the standard errors are clustered \citep{wooldridge2021}. However, it is important to note that when using a fixed effects Poisson model, the estimators may be less precise because only the within-district variation is used for identification compared to the pooled estimator.\footnote{Observations are dropped from the model if all observations of a district are 0, i.e., if a district has no within-variation. If districts are dropped, I report the number of dropped districts in respective table or figure notes.} Therefore, inference based on statistical significance is more conservative because standard errors may be higher.
	
	Crucial for difference-in-differences specifications, inference depends on the parallel trends assumption not being violated. To test whether the parallel trends assumption holds in a nonlinear setting, I perform a cluster-robust Wald test for the joint significance of $\hat{\gamma}_{2010},\dots,\hat{\gamma}_{2014}$ as suggested by \citet{wooldridge2022} -- in addition to graphical inspection. The test statistics show whether pre-treatment period estimators are statistically different from zero. Thus, insignificance of the test statistics indicates no violation of the parallel trends assumption. If the parallel trends assumption is violated, a linear pre-trend correction is added to the conditional mean specification as follows.
	\begin{align}\label{eq:DIDPOSFE_TA}
		\begin{split}
			E(y_{s,t}|c_{s},&\textbf{x}_{s,2010},\dots,\textbf{x}_{s,2019}) = \\\
			&c_{s}exp\left(\sum_{t=2010, t\neq2014}^{2019}\delta_{t} Year_{t} + \sum_{t=2015}^{2019}\gamma_{t} Bite_{s}\cdot Year_{t} + \pi Bite_{s} \cdot Trend_{t}\right) 
		\end{split}
	\end{align}
	Here, only post-treatment periods are interacted with the minimum wage bite while controlling for a bite-specific trend estimated by $\pi$. Since, pre-treatment periods are excluded in the second term, $\pi$ captures the bite-specific pre-trend and corrects $\gamma_{2015},\dots,\gamma_{2019}$ for it. However, a linear pre-trend correction is not formally derived yet for a fixed effects Poisson model. Therefore, I check whether $\pi$ actually reflects a bite-specific linear pre-trend in a non-linear setting. I suggest estimating a model only for pre-treatment periods, including only year-dummies and $Bite_{s}\cdot Trend_{t}$ as independent variables. If the estimate of $Bite_{s} \cdot Trend_{t}$ is very similar to that of $\pi$, it suggests that correcting for a linear bite-specific pre-trend works in a non-linear setting.
	
	\section{Internal migration and reallocation between workplace districts}\label{sec:int_mig}
	
	In this section, I analyze and discuss empirical results on the potential effects of the uniform minimum wage introduction in Germany on internal migration and corresponding reallocation of individuals between workplace districts. Moreover, I show several robustness checks.
	
	\subsection{Treatment effects on internal migration}\label{sec:ana1_intmig}
	
	I start by estimating the conditional fixed effects Poisson model with the conditional mean specified as in \fref{eq:DIDposFE} on the outcome variables measuring migration flows into or out of a district. For the baseline results, I estimate the model once for the total population and separately for individuals with migrant background and native-born. 
	
	\begin{figure}[ptbh]
		\caption{Treatment effects on internal migration flows}\label{fig:baseline}
		\centering
		\includegraphics[width=.9\textwidth]{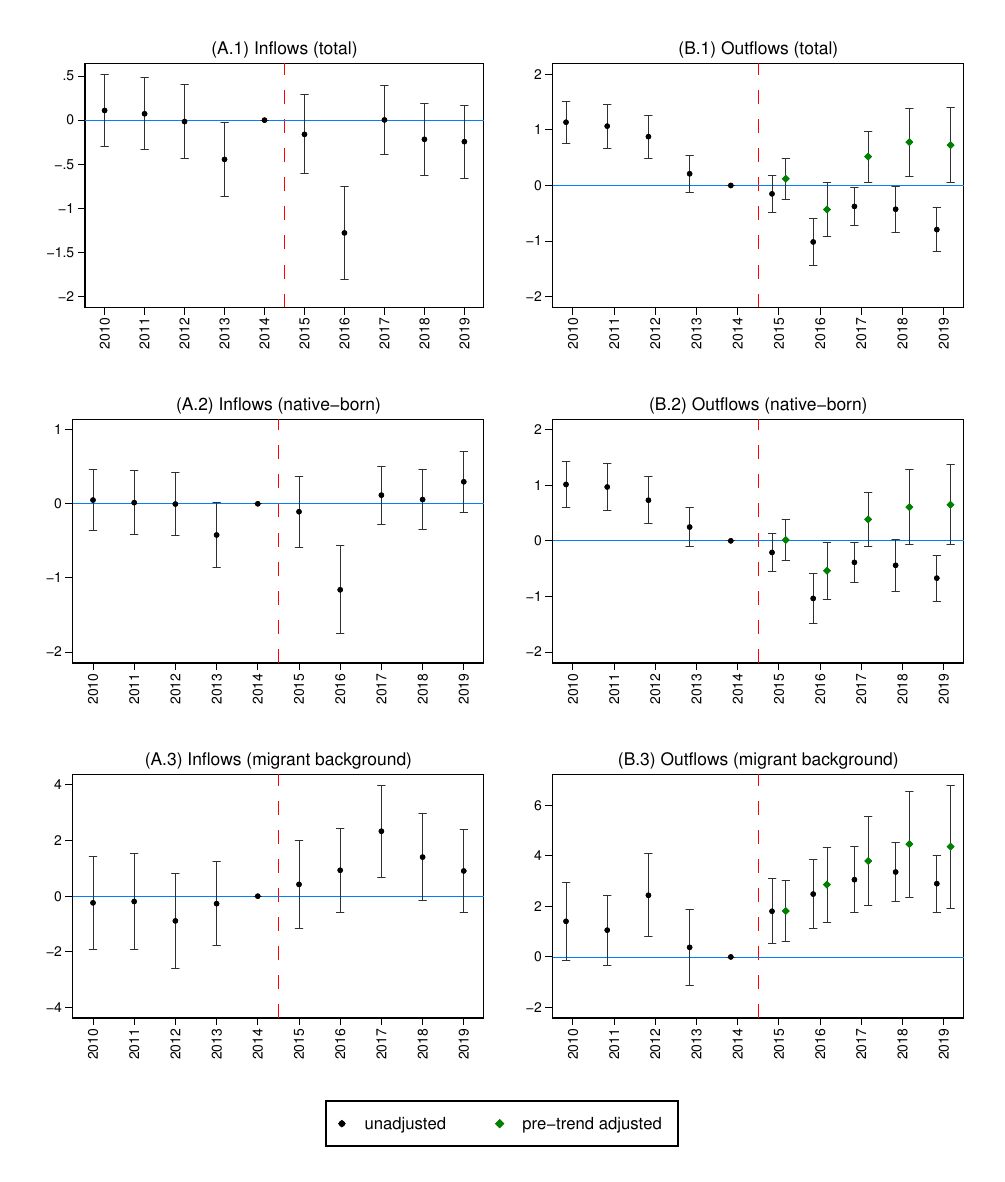} \\
		\begin{singlespace}
			\justifying \noindent \footnotesize \textit{Notes}: Cluster-robust standard errors are estimated and 95\% confidence intervals are displayed. $N=4,000$ over 400 districts. Coefficients shown are $\hat{\gamma}_{t}$ from estimating a fixed effects Poisson model with the conditional mean specified as in \fref{eq:DIDposFE}. The pre-trend Wald test suggests a significant pre-trend for following specifications: (B.1), $p= 0.000$; (B.2), $p= 0.000$; (B.3), $p= 0.024$. For the respective specifications, trend-adjusted coefficients are estimated by specifying the conditional mean as in \fref{eq:DIDPOSFE_TA}. For all pre-trend corrections, $\hat{\pi}$ and the alternative estimate for the coefficient on $Bite_{s} \cdot Trend_{t}$ are very similar -- as described in \fref{sec:ident1} to check whether \fref{eq:DIDPOSFE_TA} sufficiently corrects for a linear pre-trend. The magnitude of the average treatment effects should be interpreted by multiplying $\hat{\gamma}_{t}$ with the average treatment intensity of about 10\%. \textit{Source: SIAB, own calculations}    
		\end{singlespace}
	\end{figure}
	
	\Fref{fig:baseline} shows the respective baseline results. Graphs (A.1) to (A.3) in the left column illustrate the average treatment effects on inflows for the years 2015 to 2019, separately by row for the total population, native-born, and individuals with migrant background. In contrast, graphs (B.1) to (B.3) in the right column illustrate the treatment effects on outflows for the respective samples. The introduction of the minimum wage generally has no significant effect on individuals moving to high-bite regions. This is also the case for the native-born and migrant background sub-samples. In contrast, total out-migration increases significantly in the years 2017 to 2019 compared to the last pre-treatment period, on average by about 10\% due to the introduction of the minimum wage.\footnote{Average treatment effects are calculated as $\hat{\gamma}_{t}$ times the average treatment intensity, i.e. the average value of $Bite_{s}$. The average value is 10\% (see \fref{fig:soc_econ}).} This provides some initial evidence in favor of the main hypothesis, i.e. that out-migration increases in high-bite districts due to a uniform minimum wage introduction. Looking at the sub-samples, the effect for native-born is about the same size, but hardly significant for the years 2017 to 2019. In contrast, the introduction of the minimum wage increases the outflows of individuals with a migrant background by 30\% on average for all post-treatment periods, with an increasing tendency over time.\footnote{It is worth noting that the average treatment effects corrected for pre-trends are estimated less precisely due to less variation when including the pre-trend correction as in \fref{eq:DIDPOSFE_TA}. However, this results in higher standard errors, so the confidence intervals provide a more conservative view of the statistical significance of the estimators.} 
	
	\subsection{Robustness of the baseline results}\label{sec:robust1}
	
	\begin{figure}[tb]
		\caption{Baseline treatment effects on outflows estimated with Destatis data}\label{fig:robust_dest}
		\centering 
		\includegraphics[scale=0.23]{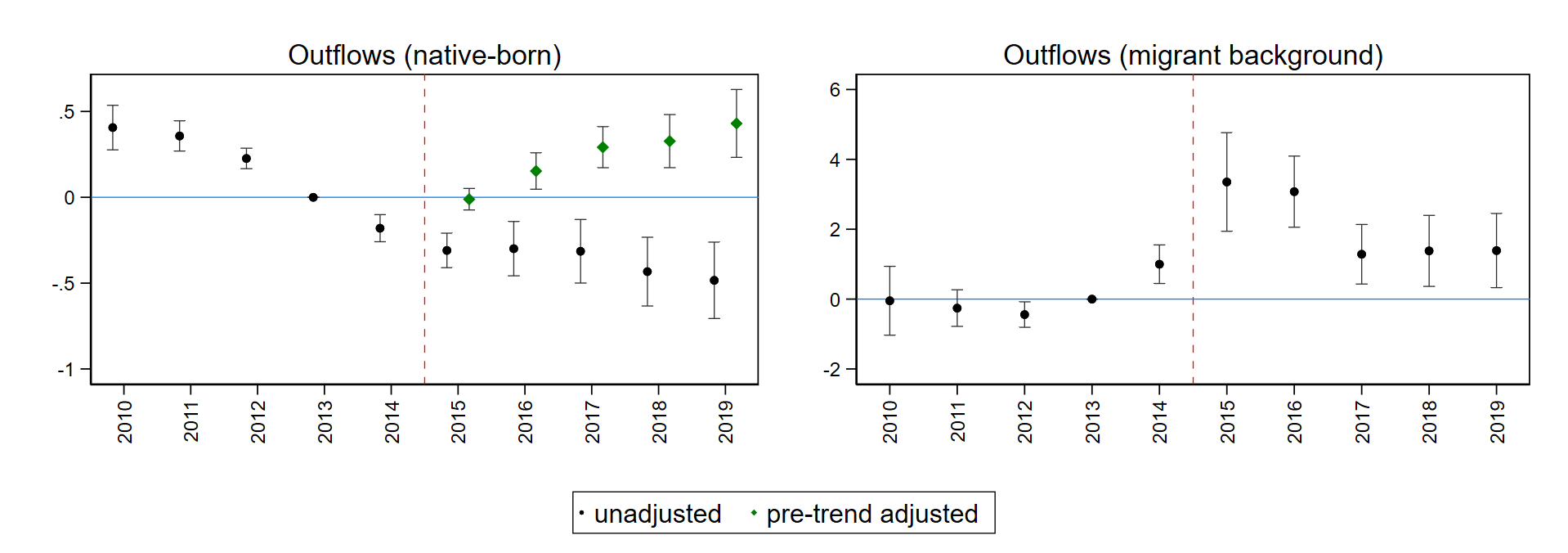}
		\begin{singlespace}
			\justifying \noindent \footnotesize \textit{Notes}: Cluster-robust standard errors are estimated and 95\% confidence intervals are displayed. $N=4,000$ over 400 districts. Coefficients shown are $\hat{\gamma}_{t}$ from estimating a fixed effects Poisson model with the conditional mean specified as in \fref{eq:DIDposFE}. The pre-trend Wald test suggests a significant pre-trend for following specification: Outflows (native-born), $p= 0.000$. For the respective specification, trend-adjusted coefficients are estimated by specifying the conditional mean as in \fref{eq:DIDPOSFE_TA}. For all pre-trend corrections, $\hat{\pi}$ and the alternative estimate for the coefficient on $Bite_{s} \cdot Trend_{t}$ are very similar -- as described in \fref{sec:ident1} to check whether \fref{eq:DIDPOSFE_TA} sufficiently corrects for a linear pre-trend. The magnitude of the average treatment effects should be interpreted by multiplying $\hat{\gamma}_{t}$ with the average treatment intensity of about 10\%. As a robustness check the results should be compared to the baseline estimates on outflows presented in \fref{fig:baseline}. \textit{Source: Regionaldatenbank (2024b), own calculations}
		\end{singlespace}
	\end{figure}
	
	To show the validity of my baseline results I perform several robustness checks. First, I re-estimate my baseline results for the outflows of both groups using full population internal migration data from the German Federal Statistical Office (Destatis) (Regionaldatenbank, 2024b). The results are reported in \fref{fig:robust_dest}. The treatment effects for both groups have roughly the same size as the estimates reported in \fref{fig:baseline} and are significant, further confirming the robustness in measuring flows by aggregating the SIAB. However, as the Destatis data collects civil register data\footnote{Destatis aggregates data from civil registers organized at the municipality level.}, there is no information on the qualification of individuals, nor information on wages or occupational groups. Therefore, a detailed analysis of the groups affected by the introduction of the minimum wage, as I have done with the SIAB, would not be possible. Moreover, the comparability of migration flows after 2016 with earlier periods is limited due to methodological changes, which may pose an additional threat to identification, although the results on outflows seem reliable.
	
	\begin{table}[tbh]
		\centering
		\caption{Robustness check: Estimates on outflows including marginal employment}\label{tab:robust_margin}
		\begin{tabularx}{\textwidth}{l*{4}{D{.}{.}{3.2}}}
			\toprule
			& \multicolumn{2}{c}{migrant background} & \multicolumn{2}{c}{native-born} \\
			\cmidrule(lr){2-3} \cmidrule(lr){4-5}
			& \multicolumn{1}{c}{w/o marg. empl.} & \multicolumn{1}{c}{w/ marg. empl.} & \multicolumn{1}{c}{w/o marg. empl.} & \multicolumn{1}{c}{w/ marg. empl.} \\
			\midrule
			$\hat{\gamma}_{2010}$ & 1.41^{*} & 1.01 & 1.01^{***} & 1.10^{***} \\
			$\hat{\gamma}_{2011}$ & 1.06 & 0.71 & 0.97^{***} & 1.00^{***} \\
			$\hat{\gamma}_{2012}$ & 2.44^{***} & 1.64^{**} & 0.73^{***} & 0.74^{***} \\
			$\hat{\gamma}_{2013}$ & 0.38 & -0.06 & 0.25 & 0.25 \\
			$\hat{\gamma}_{2014}$ & & & &  \\
			$\hat{\gamma}_{2015}$ & 1.81^{***} & 0.84 & -0.21 & -0.12 \\
			$\hat{\gamma}_{2016}$ & 2.49^{***} & 2.01^{***} & -1.04^{***} & -0.98^{***} \\
			$\hat{\gamma}_{2017}$ & 3.06^{***} & 2.35^{***} & -0.39^{**} & -0.34^{*} \\
			$\hat{\gamma}_{2018}$ & 3.36^{***} & 2.68^{***} & -0.44^{*} & -0.33 \\
			$\hat{\gamma}_{2019}$ & 2.9^{***} & 2.32^{***} & -0.67^{***} & -0.58^{***} \\
			\bottomrule
		\end{tabularx}
		\vspace{1mm}
		\begin{singlespace}
			\justifying \noindent \footnotesize \textit{Notes}: Estimates for $\gamma_t$ are provided for the baseline samples without marginal employment (w/o marg. empl.) and as a sensitivity analysis with marginal employment (w/ marg. empl.). $\gamma_t$ are estimated with a conditional fixed effects Poisson model with the conditional mean specified as in \fref{eq:DIDposFE}. $\hat{\gamma}_{2014}$ is the base category. Cluster robust standard errors are estimated. $^{*} p < 0.10$, $^{**} p < 0.05$, $^{***} p < 0.01$.  \textit{Source: SIAB, own calculations}
		\end{singlespace}
	\end{table}
	
	Second, I re-estimate the baseline results by including marginally employed workers, since they are likely to be one of the groups most affected by the minimum wage introduction.\footnote{I only report and compare non-pre-trend adjusted estimates in \fref{tab:robust_margin} and \fref{tab:robust_specs}, since I am only interested in the magnitude and direction of the results.} \Fref{tab:robust_margin} reports the respective estimates compared to the estimates excluding marginally employed workers. For individuals with a migrant background, the estimates for $\gamma_t$ including marginally employed are slightly lower than those excluding marginally employed. This is consistent with the assumption that marginally employed workers are unlikely to move due to the introduction of the minimum wage. For native-born, the estimates are fairly similar in size and significance. Both provide further support for the robustness of my results. 
	
	\begin{table}[htbp]
		\centering
		\caption{Robustness check: Estimates on outflows varying between count data models}\label{tab:robust_specs}%
		\begin{tabularx}{\textwidth}{l*{6}{D{.}{.}{3.2}}}
			\toprule
			& \multicolumn{3}{c}{migrant background} & \multicolumn{3}{c}{native-born} \\
			\cmidrule(lr){2-4} \cmidrule(lr){5-7}
			& \multicolumn{1}{c}{(1)} & \multicolumn{1}{c}{(2)} & \multicolumn{1}{c}{(3)} & \multicolumn{1}{c}{(4)} & \multicolumn{1}{c}{(5)} & \multicolumn{1}{c}{(6)} \\
			& \multicolumn{1}{c}{FE Poisson} & \multicolumn{1}{c}{pooled Poisson} & \multicolumn{1}{c}{log-linear} & \multicolumn{1}{c}{FE Poisson} & \multicolumn{1}{c}{pooled Poisson} & \multicolumn{1}{c}{log-linear} \\
			\midrule
			$\hat{\gamma}_{2010}$ & 1.41^{*} & 1.30^{*}  & 22.22^{*} & 1.01^{***} & 0.95^{***} & 1.00^{***} \\
			$\hat{\gamma}_{2011}$ & 1.06  & 0.98  & 10.59 & 0.97^{***} & 0.91^{***} & 0.95^{***} \\
			$\hat{\gamma}_{2012}$ & 2.44^{***} & 2.27^{***} & 26.61^{***} & 0.73^{***} & 0.68^{***} & 0.66^{**} \\
			$\hat{\gamma}_{2013}$ & 0.38  & 0.35  & 32.08^{***} & 0.25  & 0.23  & 0.29 \\
			$\hat{\gamma}_{2014}$ &  &  &  &  &  &  \\
			$\hat{\gamma}_{2015}$ & 1.81^{***} & 1.68^{***} & 23.58^{***} & -0.21 & -0.20  & -0.20 \\
			$\hat{\gamma}_{2016}$ & 2.49^{***} & 2.32^{***} & 38.02^{***} & -1.04^{***} & -0.96^{***} & -1.1^{***} \\
			$\hat{\gamma}_{2017}$ & 3.06^{***} & 2.86^{***} & 36.95^{***} & -0.39^{**} & -0.36^{**} & -0.41^{*} \\
			$\hat{\gamma}_{2018}$ & 3.36^{***} & 3.15^{***} & 41.92^{***} & -0.44^{*} & -0.41^{*} & -0.40 \\
			$\hat{\gamma}_{2019}$ & 2.90^{***} & 2.71^{***} & 46.87^{***} & -0.67^{***} & -0.62^{***} & -0.83^{***} \\
			\bottomrule
		\end{tabularx}
		\vspace{1mm}
		\begin{singlespace}
			\justifying \noindent \footnotesize \textit{Notes}: Estimates for $\gamma_t$ are provided for the baseline samples. $\hat{\gamma}_{t}$ are estimated in column (1) and (4) with a conditional fixed effects Poisson model with the conditional mean specified as in \fref{eq:DIDposFE}. In column (2) and (5) $\hat{\gamma}_{t}$ are estimated with a pooled Poisson model with the conditional mean specified as in \fref{eq:DIDpos}. In column (3) and (6) $\hat{\gamma}_{t}$ is estimated with a log-linear model as in \fref{eq:loglin}. $\hat{\gamma}_{2014}$ is the base category. Cluster robust standard errors are estimated. $^{*} p < 0.10$, $^{**} p < 0.05$, $^{***} p < 0.01$.  \textit{Source: SIAB, own calculations}
		\end{singlespace}
	\end{table}
	
	Third, I compare the baseline results with two alternative specifications for count data, namely I estimate the difference-in-differences specification with a pooled Poisson model (see \fref{eq:DIDpos}) and with a log-linear fixed effects model. For the log-linear fixed effects model the estimator takes the following form.
	\begin{align}\label{eq:loglin}
		ln(y_{s,t}) =  c_{s} + \sum_{t=2010, t\neq2014}^{2019}\delta_{t} Year_{t} + \sum_{t=2010, t\neq2014}^{2019}\gamma_{t} Bite_{s}\cdot Year_{t}+ u_{s,t}  
	\end{align}
	Since the log-linear case drops zeros in the dependent variable, I add a very small number to all zero values before taking the natural logarithm, which is common practice when using the corresponding model with count data. \Fref{tab:robust_specs} provides the corresponding estimates. For the pooled Poisson model, the results are similar in magnitude and significance. This further supports the robustness of my main results. The estimates for the log-linear model are similar in size and significance for the native-born sub-sample which does not include zeros. For the migrant background sub-sample, the log-linear model estimates have the same direction but are about ten times higher than the conditional fixed effects Poisson estimates. This is probably due to the presence of zeros in the dependent variable for the migrant background sub-sample. Here, the estimator of the log-linear model is inconsistent and upward biased once the dependent variable includes zeros. Moreover, the results depend on the very small number that is added to the zeros to obtain the estimates. Since the estimates are upward biased, it appears that the linear estimator puts too much weight on changes in the dependent variable from zero to any positive integer in high-bite districts due to the minimum wage introduction. Thus, the log-linear model provides further support for the robustness of my results in terms of effect direction and highlights the conditional fixed-effects Poisson estimator as the preferred econometric model, especially when the dependent variable contains zeros.
	
	\subsection{Heteregenous treatment effects: Employment status and regional variation}\label{sec:ana1_hettreat}
	
	Looking at the total population may provide a disturbed picture because the samples include individuals who are unlikely to have been subject to the introduction of the minimum wage. Therefore, I re-estimate the model by employment status, namely unemployment, low-skilled employment, and medium-/high-skilled employment.\footnote{The SIAB includes only daily wages and no measure of hours worked. Therefore, I consider low-skilled work as a proxy for being subject to the minimum wage.} Here, unemployed and low-skilled workers are likely to be most affected by the introduction of the minimum wage, as theoretically derived in \fref{sec:t_unemp} and \fref{sec:t_migbac}. The sample of medium-/high-skilled workers serves as a quasi-placebo group, since the internal migration of medium-/high-skilled workers should theoretically not be affected by the minimum wage introduction, since their wages were not affected.\footnote{For comparison, see e.g. figure 8 in \citet{bossler2023}. They show that only wages at the 50\% percentile and below were affected by the introduction of the minimum wage, while no effect can be found for wages at higher percentiles. Wages at higher percentiles most likely correspond to wages for medium-/high-skilled workers, while wages at lower percentiles correspond to wages for low-skilled workers.}
	
	\begin{figure}[ptbh]
		\caption{Treatment effects on outflows by employment status}\label{fig:het_employment}
		\centering
		\includegraphics[width=.9\textwidth]{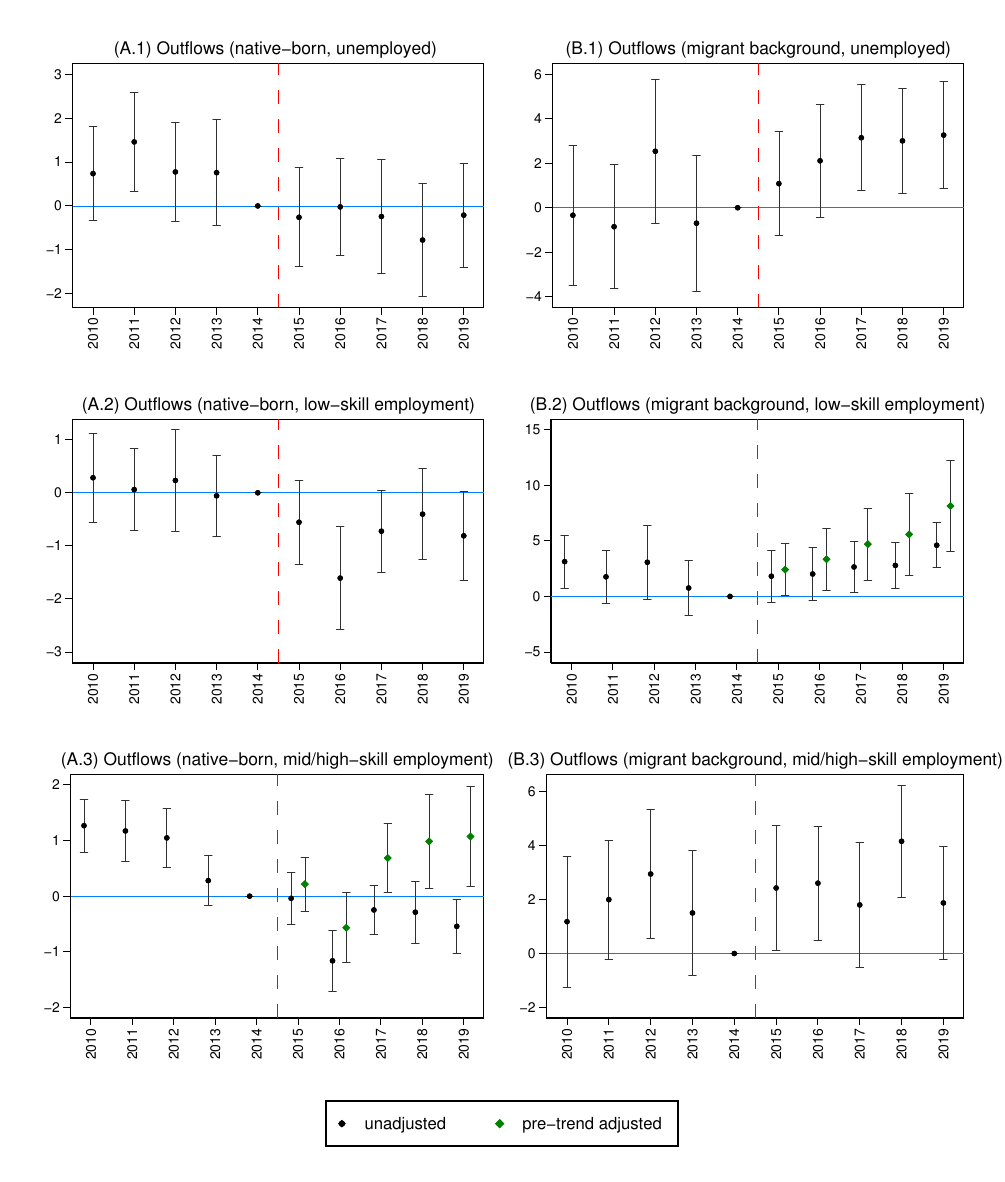} \\
		\begin{singlespace}
			\justifying \noindent \footnotesize \textit{Notes}: Cluster-robust standard errors are estimated and 95\% confidence intervals are displayed. $N=4,000$ over 400 districts.  For specifications (B.2), 1 district and (B.3), 2 districts are omitted due to all zero outcomes in the dependent variable. Coefficients shown are $\hat{\gamma}_{t}$ from estimating a fixed effects Poisson model with the conditional mean specified as in \fref{eq:DIDposFE}. The pre-trend Wald test suggests a significant pre-trend for following specifications: (A.3), $p= 0.000$; (B.2), $p= 0.096$. For the respective specifications, trend-adjusted coefficients are estimated by specifying the conditional mean as in \fref{eq:DIDPOSFE_TA}. For all pre-trend corrections, $\hat{\pi}$ and the alternative estimate for the coefficient on $Bite_{s} \cdot Trend_{t}$ are very similar -- as described in \fref{sec:ident1} to check whether \fref{eq:DIDPOSFE_TA} sufficiently corrects for a linear pre-trend. The magnitude of the average treatment effects should be interpreted by multiplying $\hat{\gamma}_{t}$ with the average treatment intensity of about 10\%. \textit{Source: SIAB, own calculations}
		\end{singlespace}
	\end{figure}
	
	\Fref{fig:het_employment} illustrates the average treatment effects on outflows by employment status, where the left column graphs show estimates for the native-born and the right column graphs show estimates for individuals with a migrant background. For the native-born, the hardly significant positive treatment effect on outflows diminishes after estimating the model separately for unemployed and low-skilled workers. This suggests that the introduction of the minimum wage has no effect on internal migration for both groups. For individuals with a migrant background in low-skilled employment, the pre-trend corrected average treatment effects are initially very similar in magnitude to the overall average effect, showing a significant increase in average outflows from high-bite districts of around 25\% due to the introduction of the minimum wage. Interestingly, the average treatment effect increases over time, nearly tripling in size by 2019. Furthermore, I also find an effect of the minimum wage introduction on the number of outflows of unemployed individuals with a migrant background, but only for the years after 2017. Here, the respective average outflows increase significantly by about 30\% due to the introduction of the minimum wage, suggesting that unemployed individuals with a migrant background seem to follow the migration decision of low-skilled workers with some delay. Both results for individuals with a migrant background are further supported by insignificant estimates of $\gamma_t$ for the quasi-placebo sample of medium-/high-skilled workers.\footnote{\Fref{tab:robust_margin_ls} to \fref{tab:robust_specs_unemp} in \fref{app:apx_robust} provide robustness checks as in \fref{sec:robust1} for the low-skilled and unemployed sub-samples which further supports the validity of the results.}
	
	Overall, only individuals with a migrant background react to the introduction of the minimum wage by moving out of high-bite districts, while native-born individuals tend to stay in their districts. This difference may be due to stronger personal ties of native-born individuals to their home region, e.g. by giving more weight to factors such as proximity to family and friends than to perceived wages and unemployment rates when evaluating utility. On the other hand, individuals with a migrant background seem to be more flexible in their decision to move, as they are less likely to be tied to a district, e.g. because their family lives abroad, lives in cities outside of high-bite rural regions or is scattered across Germany. Therefore, individuals with a migrant background are more likely to take advantage of changing economic conditions by giving more weight to economic incentives by evaluating their utility. This result is consistent with the findings of \citet{cadena2014}, which suggest that individuals with migrant background are more sensitive to changing labor market conditions compared to natives. Relating these results to the theoretical context, the initial out-migration of low-skilled workers is most likely driven by an relatively higher increase in perceived unemployment rates in high-bite districts than in low-bite districts. Subsequently, the out-migration of low-skilled workers may increase perceived (short-term) unemployment rates in high-bite districts and decrease (short-term) unemployment rates in the district to which they migrate -- if they find new employment. Therefore, the destination may become relatively more attractive over time, i.e. the indirect utility for an individual in the destination becomes higher, while the indirect utility perceived in high-bite districts becomes lower. This would explain, on the one hand, the increasing tendency of out-migration of low-skilled workers with a migrant background over time and, on the other hand, the effect on the unemployed. Due to the out-migration of the unemployed, the increase in perceived differences in unemployment rates may become lower. Another explanation for the increasing out-migration of low-skilled workers could be network effects between individuals with a migrant background. For example, some may find a new occupation after migrating from a high-bite district. Within the community, information is spread about better employment opportunities following the change in economic conditions, thus they follow their peers. Also, unemployed individuals may follow the migration decision of partners working in low-skilled occupations who migrate due to the introduction of the minimum wage. However, due to missing information on the relationships between individuals in the SIAB, it is impossible to decompose the effect, leaving the decomposition for future research.
	
	Given the findings on the migration of individuals with a migrant background, it is interesting to analyze to which regions they migrate after the introduction of the minimum wage. To shed some light on this, I separately estimate my main specification by switching between dependent variables that distinguish between outflows within or across certain regional levels. Here, I distinguish between outflows within 150 km or beyond 150 km, outflows within or across labor market regions, and outflows within or across the former East and West German borders.     
	
	\begin{figure}[ptbh]
		\caption{Treatment effects on outflows by regional variation}\label{fig:het_dist}
		\centering
		\includegraphics[width=.9\textwidth]{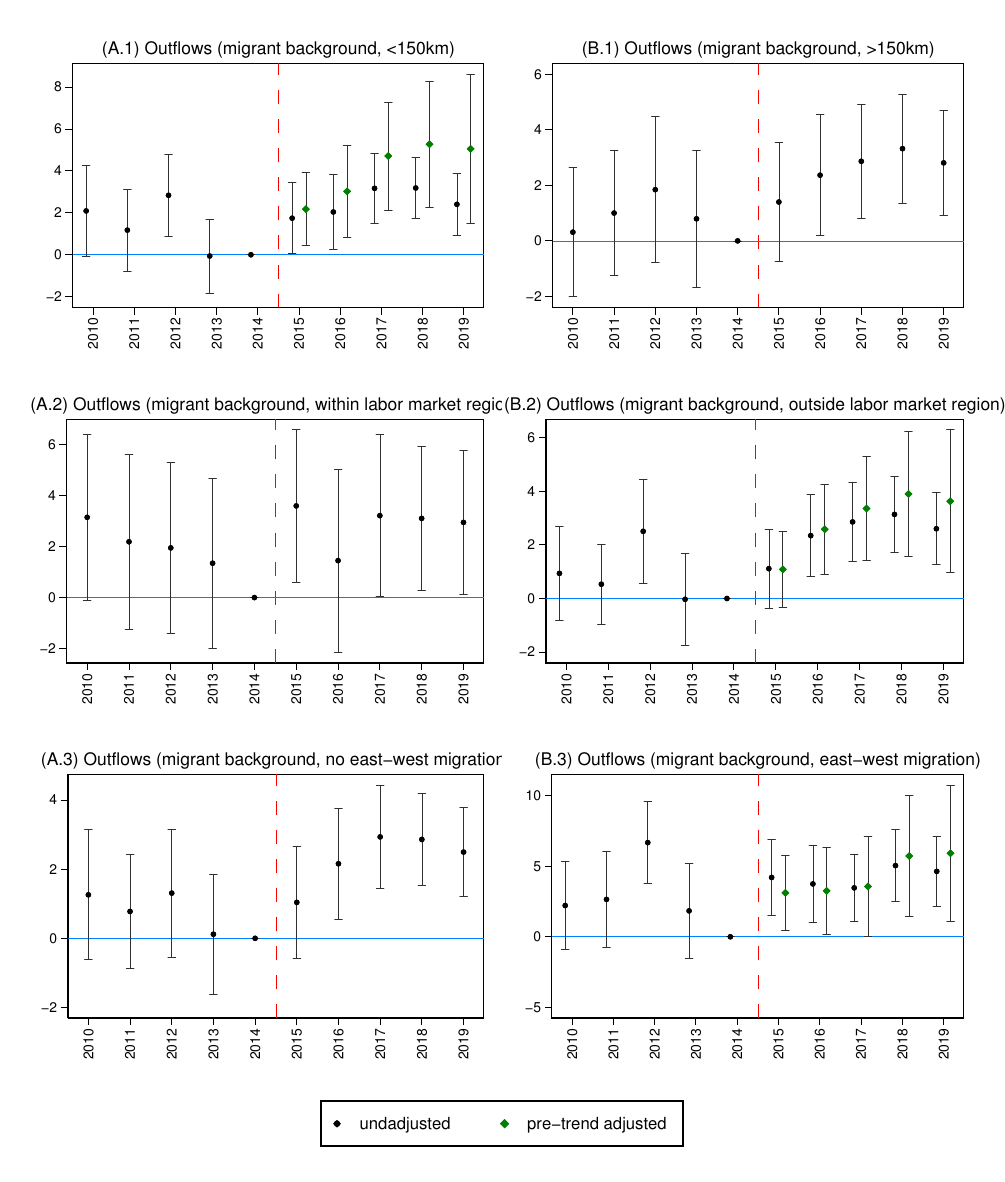} \\
		\begin{singlespace}
			\justifying \noindent \footnotesize \textit{Notes}: Cluster-robust standard errors are estimated and 95\% confidence intervals are displayed. $N=4,000$ over 400 districts. For specifications (A.2), 179 districts, (B.1), 1 district and (B.3), 5 districts are omitted due to no within-variation. Coefficients shown are $\hat{\gamma}_{t}$ from estimating a fixed effects Poisson model with the conditional mean specified as in \fref{eq:DIDposFE}. The pre-trend Wald test suggests a significant pre-trend for following specifications: (A.1), $p= 0.015$; (B.2), $p= 0.034$; (B.3), $p= 0.000$. For the respective specifications, trend-adjusted coefficients are estimated by specifying the conditional mean as in \fref{eq:DIDPOSFE_TA}. For all pre-trend corrections, $\hat{\pi}$ and the alternative estimate for the coefficient on $Bite_{s} \cdot Trend_{t}$ are very similar -- as described in \fref{sec:ident1} to check whether \fref{eq:DIDPOSFE_TA} sufficiently corrects for a linear pre-trend. The magnitude of the average treatment effects should be interpreted by multiplying $\hat{\gamma}_{t}$ with the average treatment intensity of about 10\%. \textit{Source: SIAB, own calculations}    
		\end{singlespace}
		
	\end{figure}
	
	\Fref{fig:het_dist} shows the respective average treatment effects where the first row corresponds to outflows by distance, the second row to labor market regions and the last row to former East and West German borders. Except for outflows within labor market regions\footnote{The size of the average treatment effects is about the same size as expected from the baseline results, but the standard errors are very high, since the treatment effect is very imprecisely estimated due to 179 districts (1790 Observations) being dropped because of all zero outflows.}, all graphs show significant average treatment effects after 2016. Distance does not seem to play an important role, as the average treatment effects are very similar in size. Interestingly, the change of labor market regions seems to be an important aspect for individuals with a migrant background when moving due to the introduction of the minimum wage, as average outflows across labor market regions increase by about 30 to 40\% in high-bite districts. Average outflows across former East and West German borders also increase initially by around 30\% and become even higher in 2018 and 2019, i.e. average outflows from high-bite districts in 2018 and 2019 increase by around 50\% due to the introduction of the minimum wage. Within the boundaries, average outflows also increase, but only by about 20\% to 30\% after 2016. As high-bite districts are concentrated within former East German borders, it is apparent that individuals with a migrant background tend to leave economically weaker former East German regions and thus leave weaker labor market regions for (still) economically stronger West German districts where more larger establishments are located. This further supports the view that individuals with a migrant background are more likely to take advantage of changing economic conditions, as they tend to relocate to economically stronger regions. For native-born the effect remains similar in size but insignificant in terms of regional variation (refer to \fref{fig:het_dist_mig0_apx} in \fref{app:apx_res}).
	
	\subsection{Treatment effects on changes of the workplace district}\label{sec:ana1_chgao}
	
	After examining the effect of the German minimum wage introduction in 2015 on internal migration, I will analyze whether out-migration from high-bite districts also corresponds to a change in the district in which individuals work. A change in the district of work likely reflects a reallocation across establishments, similar to \citet{dustmann2021}. Individuals could also respond to changing economic conditions by simply changing their district of work while residing in their home district. I do so by estimating the conditional fixed effects Poisson model with the conditional mean specified as in \fref{eq:DIDposFE} -- in the case of a linear pre-trend as in \fref{eq:DIDPOSFE_TA} -- on the number of individuals who change their workplace out of a district. As in \fref{sec:ana1_hettreat}, I run the analysis separately by skill level of employment, with the sub-sample of medium-/high-skilled employment again serving as a quasi-placebo group to underscore the robustness of the results for low-skill employment.
	
	\begin{figure}[tbh]
		\caption{Treatment effects on changing the district of work}\label{fig:att_ao_chg}
		\centering
		\includegraphics[width=0.89\textwidth]{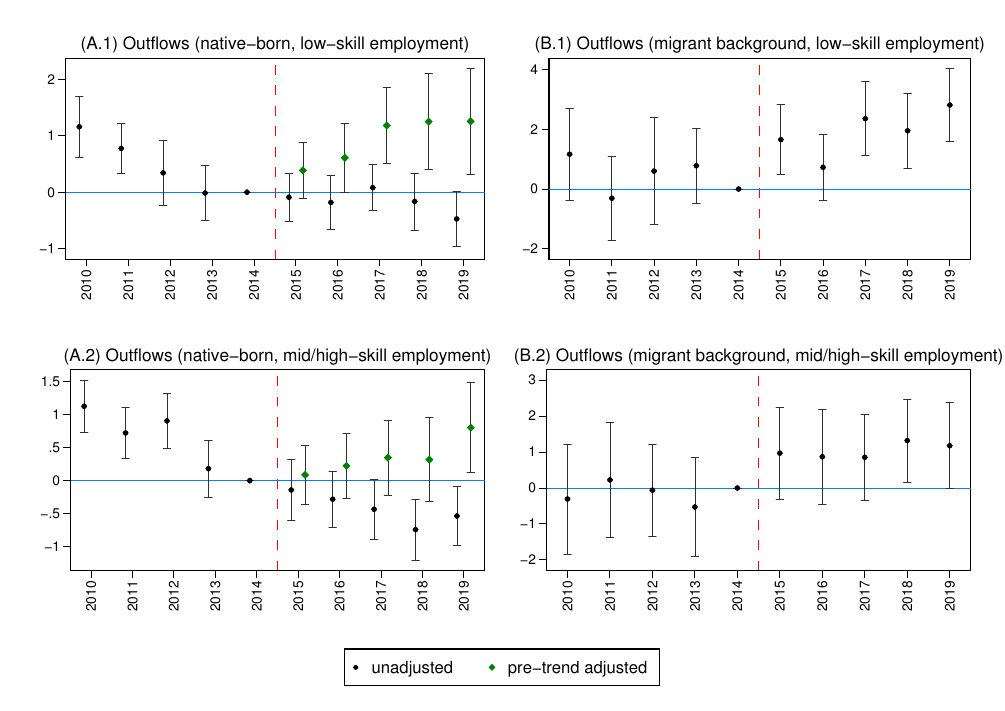} \\
		\begin{singlespace}
			\justifying \noindent \footnotesize \textit{Notes}: Cluster-robust standard errors are estimated and 95\% confidence intervals are displayed. $N=4,000$ over 400 districts. Coefficients shown are $\hat{\gamma}_{t}$ from estimating a fixed effects Poisson model with the conditional mean specified as in \fref{eq:DIDposFE}. The pre-trend Wald test suggests a significant pre-trend for following specifications: (A.1), $p= 0.000$; (A.2), $p= 0.000$. For the respective specifications, trend-adjusted coefficients are estimated by specifying the conditional mean as in \fref{eq:DIDPOSFE_TA}. For all pre-trend corrections, $\hat{\pi}$ and the alternative estimate for the coefficient on $Bite_{s} \cdot Trend_{t}$ are very similar -- as described in \fref{sec:ident1} to check whether \fref{eq:DIDPOSFE_TA} sufficiently corrects for a linear pre-trend. The magnitude of the average treatment effects should be interpreted by multiplying $\hat{\gamma}_{t}$ with the average treatment intensity of about 10\%. \textit{Source: SIAB, own calculations}
		\end{singlespace}
	\end{figure}
	
	\Fref{fig:att_ao_chg} shows respective average treatment effects on the number of individuals changing their workplace out of a district. The graphs in the left column illustrate estimates for native-born workers and the graphs in the right column illustrate estimates for workers with a migrant background. Workers with a migrant background in low-skilled employment relocate their workplace on average by 20\% to 30\% out of high-bite districts due the introduction of the minimum wage. This is consistent with their decision to move, as they tend to move across East-West German borders and across labor market regions, which likely corresponds to a change of the employer when moving. Also, native-born low-skilled workers change their workplace away from high-bite districts on average by about 10\% for the periods 2017 to 2019 due to the minimum wage introduction. In contrast to individuals with a migrant background, I find no effect on the migration flows of low-skilled native-born workers in \fref{sec:ana1_hettreat}. Thus, low-skilled native-born workers also respond to changing local labor market conditions due to the introduction of the minimum wage, but only by reallocating across establishments away from high-bite districts while residing in their home district. This is consistent with the findings of \citet{dustmann2021}, who found an increase in cross-municipality commuting due to the introduction of the minimum wage. I further support these findings by showing that the results hold when analyzed at the broader district level, but only for native-born. Further, the effects on the change of the workplace for both groups are probably due to the closure of small establishments and a corresponding reallocation of the workplace across districts to larger and more stable establishments \citep{dustmann2021}. However, native-born and individuals with migrant background appear to reallocate differently. Those with a migrant background tend to move to more distant establishments where commuting is not feasible, as they also change their place of residence in comparison to natives who tend to stay in their district of residence after the minimum wage introduction. There could be several reasons for the difference in reallocation behavior. On the one hand, it could be for reasons similar to those mentioned in \fref{sec:ana1_hettreat}, namely that individuals with a migrant background are more responsive to changing labor market conditions, are less tied to districts, or have stronger network effects within their communities. Another explanation could be that native-born individuals have an advantage in obtaining nearby jobs in establishments geographically close to their districts of residency. Subsequently, the labor demand of respective large establishments is already saturated by the increasing labor supply of native-born low-skilled workers. As a result, low-skilled workers with migrant background migrate because occupations are only available in large establishments located in more distant districts. This seems likely given the higher concentration of large establishments in low-bite districts, which are mainly located in former West Germany.
	
	As expected, I find no effect of the minimum wage introduction on reallocation between workplace districts for the quasi-placebo group of medium-/high-skilled workers. This holds for both the native-born and migrant-background sub-samples which further supports the robustness of my findings.
	
	\section{Conclusion}\label{sec:conclusion}
	In this paper, I examine the effect of a uniform minimum wage in Germany in 2015 on internal migration. For the sample of low-skilled workers with a migrant background, I find an increase in the average out-migration from high-bite districts of about 20\% to 40\% due to the introduction of the minimum wage. In contrast, I find no effect on out-migration for native-born low-skilled workers. This findings suggest that individuals with a migrant background are more responsive to changing labor market conditions than native-born individuals, similar to the findings of \citet{cadena2014} for the United States.  
	
	In addition, for low-skilled workers, both native-born and with migrant background, I find an increase in reallocation between establishments away from high-bite districts due to the introduction of the minimum wage, suggesting an increase in perceived unemployment in high-bite districts that does not seem to be observable at the aggregate level since individuals already found a new occupation within a year. This is similar to \citet{dustmann2021} findings on the reallocation to large establishments due to the closure of small establishments in high-bite municipalities. However, \citet{dustmann2021} also report an increase in commuting in high-bite municipalities, which my results suggest only for native-born. On the other hand, low-skilled workers with a migrant background reallocate by also moving out of high-bite districts, suggesting that these workers reallocate to more distant establishments. This is underscored by the stronger effect of individuals with a migrant background moving across the former East-West German border and across labor market regions. 
	
	Also, I find an increase in the average number of out-migration from high-bite districts for unemployed individuals with a migration background for the years 2017 to 2019. However, the channel through which unemployed individuals migrate is not entirely clear. Firstly, unemployed individuals may follow the same migration patterns as minimum wage workers if the reservation wage is lower than the minimum wage. Second, unemployed individuals may follow the migration decision of their partners who are subject to the minimum wage. The decomposition of the results is left to future research, as the SIAB does not provide information on the relationship between individuals.
	
	The findings of this paper underscores the relevance of taking the potential effects on geographical labor mobility into account when implementing and evaluating labor market policies. Moreover, policy makers should consider the heterogeneous effect on the migration decisions between native-born individuals and those with a migrant background. As my results suggest, economically already weaker rural former East German regions could face a change in the composition of workers due to the minimum wage introduction besides the already present decreasing trend in young individuals and individuals with a migrant background. Especially, migrant workers do more often work in occupations which already face labor shortage as, e.g. occupations in the health and care sector or in manual labor\footnote{For comparison see, e.g. \hyperlink{https://www.destatis.de/DE/Presse/Pressemitteilungen/2024/03/PD24_081_125.html}{https://www.destatis.de/DE/Presse/Pressemitteilungen/2024/03/PD24\_081\_125.html}.}. Not considering heterogeneous policy effects on the migration decision of workers, could pose an additional problem for highly affected regions, i.e. a policy-induced increasing future shortage of (infrastructure-relevant) labor. Therefore, implementing policies offsetting the out-migration effect on low-skilled workers with migrant background could be beneficial for highly affected regions in the long-run.
	
	
	\clearpage
	\bibliography{sources}

\begin{thebibliography}{}

\bibitem [\protect \citeauthoryear {%
Bauer%
, Rulff%
\BCBL {}\ \BBA {} Tamminga%
}{%
Bauer%
\ \protect \BOthers {.}}{%
{\protect \APACyear {2019}}%
}]{%
bauer2019}
\APACinsertmetastar {%
bauer2019}%
\begin{APACrefauthors}%
Bauer, T\BPBI K.%
, Rulff, C.%
\BCBL {}\ \BBA {} Tamminga, M\BPBI M.%
\end{APACrefauthors}%
\unskip\
\newblock
\APACrefYearMonthDay{2019}{}{}.
\newblock
\APACrefbtitle {Berlin calling - Internal migration in Germany} {Berlin calling
  - internal migration in germany}\ \APACbVolEdTR {}{Ruhr Economic Papers\
  \BNUM~823}.
\newblock
\begin{APACrefDOI} \doi{10.4419/86788956} \end{APACrefDOI}
\PrintBackRefs{\CurrentBib}

\bibitem [\protect \citeauthoryear {%
Bhaskar%
, Manning%
\BCBL {}\ \BBA {} To%
}{%
Bhaskar%
\ \protect \BOthers {.}}{%
{\protect \APACyear {2002}}%
}]{%
bhaskar2002}
\APACinsertmetastar {%
bhaskar2002}%
\begin{APACrefauthors}%
Bhaskar, V.%
, Manning, A.%
\BCBL {}\ \BBA {} To, T.%
\end{APACrefauthors}%
\unskip\
\newblock
\APACrefYearMonthDay{2002}{}{}.
\newblock
{\BBOQ}\APACrefatitle {Oligopsony and Monopsonistic Competition in Labor
  Markets} {Oligopsony and monopsonistic competition in labor markets}.{\BBCQ}
\newblock
\APACjournalVolNumPages{Journal of Economic Perspectives}{16}{2}{155-174}.
\newblock
\begin{APACrefURL}
  \url{https://www.aeaweb.org/articles?id=10.1257/0895330027300}
  \end{APACrefURL}
\newblock
\begin{APACrefDOI} \doi{10.1257/0895330027300} \end{APACrefDOI}
\PrintBackRefs{\CurrentBib}

\bibitem [\protect \citeauthoryear {%
Bossler%
, Liang%
\BCBL {}\ \BBA {} Schank%
}{%
Bossler%
\ \protect \BOthers {.}}{%
{\protect \APACyear {2024}}%
}]{%
bossler2024}
\APACinsertmetastar {%
bossler2024}%
\begin{APACrefauthors}%
Bossler, M.%
, Liang, Y.%
\BCBL {}\ \BBA {} Schank, T.%
\end{APACrefauthors}%
\unskip\
\newblock
\APACrefYearMonthDay{2024}{}{}.
\newblock
{\BBOQ}\APACrefatitle {The Devil is in the Details: Heterogeneous Effects of
  the German Minimum Wage on Working Hours and Minijobs} {The devil is in the
  details: Heterogeneous effects of the german minimum wage on working hours
  and minijobs}.{\BBCQ}
\newblock
\APACjournalVolNumPages{arXiv preprint arXiv:2403.17206}{}{}{}.
\PrintBackRefs{\CurrentBib}

\bibitem [\protect \citeauthoryear {%
Bossler%
\ \BBA {} Schank%
}{%
Bossler%
\ \BBA {} Schank%
}{%
{\protect \APACyear {2023}}%
}]{%
bossler2023}
\APACinsertmetastar {%
bossler2023}%
\begin{APACrefauthors}%
Bossler, M.%
\BCBT {}\ \BBA {} Schank, T.%
\end{APACrefauthors}%
\unskip\
\newblock
\APACrefYearMonthDay{2023}{}{}.
\newblock
{\BBOQ}\APACrefatitle {Wage inequality in Germany after the minimum wage
  introduction} {Wage inequality in germany after the minimum wage
  introduction}.{\BBCQ}
\newblock
\APACjournalVolNumPages{Journal of Labor Economics}{41}{3}{000--000}.
\newblock
\begin{APACrefDOI} \doi{10.1086/720391} \end{APACrefDOI}
\PrintBackRefs{\CurrentBib}

\bibitem [\protect \citeauthoryear {%
Cadena%
}{%
Cadena%
}{%
{\protect \APACyear {2013}}%
}]{%
cadena2014}
\APACinsertmetastar {%
cadena2014}%
\begin{APACrefauthors}%
Cadena, B\BPBI C.%
\end{APACrefauthors}%
\unskip\
\newblock
\APACrefYearMonthDay{2013}{}{}.
\newblock
{\BBOQ}\APACrefatitle {Recent immigrants as labor market arbitrageurs: Evidence
  from the minimum wage} {Recent immigrants as labor market arbitrageurs:
  Evidence from the minimum wage}.{\BBCQ}
\newblock
\APACjournalVolNumPages{Journal of Urban Economics}{80}{}{1-12}.
\newblock
\begin{APACrefDOI} \doi{10.1016/j.jue.2013.10.002} \end{APACrefDOI}
\PrintBackRefs{\CurrentBib}

\bibitem [\protect \citeauthoryear {%
Card%
}{%
Card%
}{%
{\protect \APACyear {1992}}%
}]{%
card1992}
\APACinsertmetastar {%
card1992}%
\begin{APACrefauthors}%
Card, D.%
\end{APACrefauthors}%
\unskip\
\newblock
\APACrefYearMonthDay{1992}{}{}.
\newblock
{\BBOQ}\APACrefatitle {Using regional variation in wages to measure the effects
  of the federal minimum wage} {Using regional variation in wages to measure
  the effects of the federal minimum wage}.{\BBCQ}
\newblock
\APACjournalVolNumPages{Ilr Review}{46}{1}{22--37}.
\newblock
\begin{APACrefDOI} \doi{10.2307/2524736} \end{APACrefDOI}
\PrintBackRefs{\CurrentBib}

\bibitem [\protect \citeauthoryear {%
Card%
, Heining%
\BCBL {}\ \BBA {} Kline%
}{%
Card%
\ \protect \BOthers {.}}{%
{\protect \APACyear {2013}}%
}]{%
card2013}
\APACinsertmetastar {%
card2013}%
\begin{APACrefauthors}%
Card, D.%
, Heining, J.%
\BCBL {}\ \BBA {} Kline, P.%
\end{APACrefauthors}%
\unskip\
\newblock
\APACrefYearMonthDay{2013}{}{}.
\newblock
{\BBOQ}\APACrefatitle {Workplace heterogeneity and the rise of West German wage
  inequality} {Workplace heterogeneity and the rise of west german wage
  inequality}.{\BBCQ}
\newblock
\APACjournalVolNumPages{The Quarterly journal of economics}{128}{3}{967--1015}.
\newblock
\begin{APACrefDOI} \doi{10.1093/qje/qjt006} \end{APACrefDOI}
\PrintBackRefs{\CurrentBib}

\bibitem [\protect \citeauthoryear {%
Dustmann%
, Lindner%
, Sch\"onberg%
, Umkehrer%
\BCBL {}\ \BBA {} vom Berge%
}{%
Dustmann%
\ \protect \BOthers {.}}{%
{\protect \APACyear {2021}}%
}]{%
dustmann2021}
\APACinsertmetastar {%
dustmann2021}%
\begin{APACrefauthors}%
Dustmann, C.%
, Lindner, A.%
, Sch\"onberg, U.%
, Umkehrer, M.%
\BCBL {}\ \BBA {} vom Berge, P.%
\end{APACrefauthors}%
\unskip\
\newblock
\APACrefYearMonthDay{2021}{}{}.
\newblock
{\BBOQ}\APACrefatitle {Reallocation effects of the minimum wage} {Reallocation
  effects of the minimum wage}.{\BBCQ}
\newblock
\APACjournalVolNumPages{The Quarterly Journal of Economics}{137}{1}{267-328}.
\newblock
\begin{APACrefDOI} \doi{10.1093/qje/qjab028} \end{APACrefDOI}
\PrintBackRefs{\CurrentBib}

\bibitem [\protect \citeauthoryear {%
Dustmann%
, Ludsteck%
\BCBL {}\ \BBA {} Sch{\"o}nberg%
}{%
Dustmann%
\ \protect \BOthers {.}}{%
{\protect \APACyear {2009}}%
}]{%
dustmann2009}
\APACinsertmetastar {%
dustmann2009}%
\begin{APACrefauthors}%
Dustmann, C.%
, Ludsteck, J.%
\BCBL {}\ \BBA {} Sch{\"o}nberg, U.%
\end{APACrefauthors}%
\unskip\
\newblock
\APACrefYearMonthDay{2009}{}{}.
\newblock
{\BBOQ}\APACrefatitle {Revisiting the German wage structure} {Revisiting the
  german wage structure}.{\BBCQ}
\newblock
\APACjournalVolNumPages{The Quarterly journal of economics}{124}{2}{843--881}.
\newblock
\begin{APACrefDOI} \doi{10.1162/qjec.2009.124.2.843} \end{APACrefDOI}
\PrintBackRefs{\CurrentBib}

\bibitem [\protect \citeauthoryear {%
Fuchs-Sch\"undeln%
\ \BBA {} Sch\"undeln%
}{%
Fuchs-Sch\"undeln%
\ \BBA {} Sch\"undeln%
}{%
{\protect \APACyear {2009}}%
}]{%
schuendeln2009}
\APACinsertmetastar {%
schuendeln2009}%
\begin{APACrefauthors}%
Fuchs-Sch\"undeln, N.%
\BCBT {}\ \BBA {} Sch\"undeln, M.%
\end{APACrefauthors}%
\unskip\
\newblock
\APACrefYearMonthDay{2009}{}{}.
\newblock
{\BBOQ}\APACrefatitle {Who stays, who goes, who returns?} {Who stays, who goes,
  who returns?}{\BBCQ}
\newblock
\APACjournalVolNumPages{Economics of Transition}{17}{4}{703-738}.
\newblock
\begin{APACrefDOI} \doi{10.1111/j.1468-0351.2009.00373.x} \end{APACrefDOI}
\PrintBackRefs{\CurrentBib}

\bibitem [\protect \citeauthoryear {%
Haelbig%
, Mertens%
\BCBL {}\ \BBA {} M\"uller%
}{%
Haelbig%
\ \protect \BOthers {.}}{%
{\protect \APACyear {2023}}%
}]{%
haelbig2023}
\APACinsertmetastar {%
haelbig2023}%
\begin{APACrefauthors}%
Haelbig, M.%
, Mertens, M.%
\BCBL {}\ \BBA {} M\"uller, S.%
\end{APACrefauthors}%
\unskip\
\newblock
\APACrefYearMonthDay{2023}{}{}.
\newblock
\APACrefbtitle {Minimum Wages, Productivity, and Reallocation} {Minimum wages,
  productivity, and reallocation}\ \APACbVolEdTR {}{IZA Discussion Papers\
  \BNUM\ 16160}.
\newblock
\APACaddressInstitution{}{Institute of Labor Economics (IZA)}.
\newblock
\begin{APACrefURL} \url{https://ideas.repec.org/p/iza/izadps/dp16160.html}
  \end{APACrefURL}
\newblock
\begin{APACrefDOI} \doi{10.2139/ssrn.4457826} \end{APACrefDOI}
\PrintBackRefs{\CurrentBib}

\bibitem [\protect \citeauthoryear {%
Mitze%
\ \BBA {} Reinkowski%
}{%
Mitze%
\ \BBA {} Reinkowski%
}{%
{\protect \APACyear {2011}}%
}]{%
mitze2012}
\APACinsertmetastar {%
mitze2012}%
\begin{APACrefauthors}%
Mitze, T.%
\BCBT {}\ \BBA {} Reinkowski, J.%
\end{APACrefauthors}%
\unskip\
\newblock
\APACrefYearMonthDay{2011}{}{}.
\newblock
{\BBOQ}\APACrefatitle {Testing the neoclassical migration model: overall and
  age-group specific results for German regions} {Testing the neoclassical
  migration model: overall and age-group specific results for german
  regions}.{\BBCQ}
\newblock
\APACjournalVolNumPages{ZAF}{43}{}{277--297}.
\newblock
\begin{APACrefDOI} \doi{10.1007/s12651-010-0046-2} \end{APACrefDOI}
\PrintBackRefs{\CurrentBib}

\bibitem [\protect \citeauthoryear {%
Monras%
}{%
Monras%
}{%
{\protect \APACyear {2019}}%
}]{%
monras2019}
\APACinsertmetastar {%
monras2019}%
\begin{APACrefauthors}%
Monras, J.%
\end{APACrefauthors}%
\unskip\
\newblock
\APACrefYearMonthDay{2019}{}{}.
\newblock
{\BBOQ}\APACrefatitle {Minimum Wages and Spatial Equilibrium: Theory and
  Evidence} {Minimum wages and spatial equilibrium: Theory and
  evidence}.{\BBCQ}
\newblock
\APACjournalVolNumPages{Journal of Labor Economics}{37}{3}{853-904}.
\newblock
\begin{APACrefDOI} \doi{10.1086/702650} \end{APACrefDOI}
\PrintBackRefs{\CurrentBib}

\bibitem [\protect \citeauthoryear {%
Popp%
}{%
Popp%
}{%
{\protect \APACyear {2023}}%
}]{%
popp2023}
\APACinsertmetastar {%
popp2023}%
\begin{APACrefauthors}%
Popp, M.%
\end{APACrefauthors}%
\unskip\
\newblock
\APACrefYearMonthDay{2023}{}{}.
\newblock
{\BBOQ}\APACrefatitle {How elastic is labor demand? A meta-analysis for the
  German labor market} {How elastic is labor demand? a meta-analysis for the
  german labor market}.{\BBCQ}
\newblock
\APACjournalVolNumPages{Journal for Labor Market Research}{57:14}{}{1-21}.
\newblock
\begin{APACrefDOI} \doi{10.1186/s12651-023-00337-8} \end{APACrefDOI}
\PrintBackRefs{\CurrentBib}

\bibitem [\protect \citeauthoryear {%
Rosenbaum-Feldbr{\"u}gge%
, Stawarz%
\BCBL {}\ \BBA {} Sander%
}{%
Rosenbaum-Feldbr{\"u}gge%
\ \protect \BOthers {.}}{%
{\protect \APACyear {2022}}%
}]{%
rosenbaum2022}
\APACinsertmetastar {%
rosenbaum2022}%
\begin{APACrefauthors}%
Rosenbaum-Feldbr{\"u}gge, M.%
, Stawarz, N.%
\BCBL {}\ \BBA {} Sander, N.%
\end{APACrefauthors}%
\unskip\
\newblock
\APACrefYearMonthDay{2022}{}{}.
\newblock
{\BBOQ}\APACrefatitle {30 Years of East-West Migration in Germany: A Synthesis
  of the Literature and Potential Directions for Future Research} {30 years of
  east-west migration in germany: A synthesis of the literature and potential
  directions for future research}.{\BBCQ}
\newblock
\APACjournalVolNumPages{Comparative Population Studies}{47}{}{}.
\newblock
\begin{APACrefDOI} \doi{10.12765/CPoS-2022-08} \end{APACrefDOI}
\PrintBackRefs{\CurrentBib}

\bibitem [\protect \citeauthoryear {%
Schmucker%
, Seth%
\BCBL {}\ \BBA {} vom Berge%
}{%
Schmucker%
\ \protect \BOthers {.}}{%
{\protect \APACyear {2023}}%
}]{%
SIAB_DESC}
\APACinsertmetastar {%
SIAB_DESC}%
\begin{APACrefauthors}%
Schmucker, A.%
, Seth, S.%
\BCBL {}\ \BBA {} vom Berge, P.%
\end{APACrefauthors}%
\unskip\
\newblock
\APACrefYearMonthDay{2023}{}{}.
\newblock
{\BBOQ}\APACrefatitle {Sample of Integrated Labour Market Biographies (SIAB)
  1975 - 2021} {Sample of integrated labour market biographies (siab) 1975 -
  2021}.{\BBCQ}
\newblock
\APACjournalVolNumPages{FDZ Datenreport, 02/2023 (en), N\"urnberg}{}{}{}.
\newblock
\begin{APACrefDOI} \doi{10.5164/IAB.FDZD.2302.en.v1} \end{APACrefDOI}
\PrintBackRefs{\CurrentBib}

\bibitem [\protect \citeauthoryear {%
Silva%
\ \BBA {} Tenreyro%
}{%
Silva%
\ \BBA {} Tenreyro%
}{%
{\protect \APACyear {2006}}%
}]{%
silva2006log}
\APACinsertmetastar {%
silva2006log}%
\begin{APACrefauthors}%
Silva, J\BPBI S.%
\BCBT {}\ \BBA {} Tenreyro, S.%
\end{APACrefauthors}%
\unskip\
\newblock
\APACrefYearMonthDay{2006}{}{}.
\newblock
{\BBOQ}\APACrefatitle {The log of gravity} {The log of gravity}.{\BBCQ}
\newblock
\APACjournalVolNumPages{The Review of Economics and
  statistics}{88}{4}{641--658}.
\PrintBackRefs{\CurrentBib}

\bibitem [\protect \citeauthoryear {%
Stawarz%
\ \BBA {} Sander%
}{%
Stawarz%
\ \BBA {} Sander%
}{%
{\protect \APACyear {2019}}%
}]{%
stawarz2019}
\APACinsertmetastar {%
stawarz2019}%
\begin{APACrefauthors}%
Stawarz, N.%
\BCBT {}\ \BBA {} Sander, N.%
\end{APACrefauthors}%
\unskip\
\newblock
\APACrefYearMonthDay{2019}{}{}.
\newblock
{\BBOQ}\APACrefatitle {The impact of internal migration on the spatial
  distribution of population in Germany over the period 1991-2017} {The impact
  of internal migration on the spatial distribution of population in germany
  over the period 1991-2017}.{\BBCQ}
\newblock
\APACjournalVolNumPages{Comparative Population Studies}{44}{}{}.
\newblock
\begin{APACrefDOI} \doi{10.12765/CPoS-2020-06} \end{APACrefDOI}
\PrintBackRefs{\CurrentBib}

\bibitem [\protect \citeauthoryear {%
St{\"u}ber%
, Dauth%
\BCBL {}\ \BBA {} Eppelsheimer%
}{%
St{\"u}ber%
\ \protect \BOthers {.}}{%
{\protect \APACyear {2023}}%
}]{%
stuber2023guide}
\APACinsertmetastar {%
stuber2023guide}%
\begin{APACrefauthors}%
St{\"u}ber, H.%
, Dauth, W.%
\BCBL {}\ \BBA {} Eppelsheimer, J.%
\end{APACrefauthors}%
\unskip\
\newblock
\APACrefYearMonthDay{2023}{}{}.
\newblock
{\BBOQ}\APACrefatitle {A guide to preparing the sample of integrated labour
  market biographies (SIAB, version 7519 v1) for scientific analysis} {A guide
  to preparing the sample of integrated labour market biographies (siab,
  version 7519 v1) for scientific analysis}.{\BBCQ}
\newblock
\APACjournalVolNumPages{Journal for Labour Market Research}{57}{1}{1--11}.
\newblock
\begin{APACrefDOI} \doi{10.1186/s12651-023-00335-w} \end{APACrefDOI}
\PrintBackRefs{\CurrentBib}

\bibitem [\protect \citeauthoryear {%
Wooldridge%
}{%
Wooldridge%
}{%
{\protect \APACyear {1999}}%
}]{%
wooldridge1999}
\APACinsertmetastar {%
wooldridge1999}%
\begin{APACrefauthors}%
Wooldridge, J\BPBI M.%
\end{APACrefauthors}%
\unskip\
\newblock
\APACrefYearMonthDay{1999}{}{}.
\newblock
{\BBOQ}\APACrefatitle {Quasi-likelihood methods for count data}
  {Quasi-likelihood methods for count data}.{\BBCQ}
\newblock
\APACjournalVolNumPages{Handbook of applied econometrics volume 2:
  Microeconomics}{}{}{321--368}.
\newblock
\begin{APACrefDOI} \doi{10.1111/b.9780631216339.1999.00009.x} \end{APACrefDOI}
\PrintBackRefs{\CurrentBib}

\bibitem [\protect \citeauthoryear {%
Wooldridge%
}{%
Wooldridge%
}{%
{\protect \APACyear {2021}}%
}]{%
wooldridge2021}
\APACinsertmetastar {%
wooldridge2021}%
\begin{APACrefauthors}%
Wooldridge, J\BPBI M.%
\end{APACrefauthors}%
\unskip\
\newblock
\APACrefYearMonthDay{2021}{}{}.
\newblock
{\BBOQ}\APACrefatitle {Two-way fixed effects, the two-way mundlak regression,
  and difference-in-differences estimators} {Two-way fixed effects, the two-way
  mundlak regression, and difference-in-differences estimators}.{\BBCQ}
\newblock
\APACjournalVolNumPages{Available at SSRN 3906345}{}{}{}.
\newblock
\begin{APACrefDOI} \doi{10.2139/ssrn.3906345} \end{APACrefDOI}
\PrintBackRefs{\CurrentBib}

\bibitem [\protect \citeauthoryear {%
Wooldridge%
}{%
Wooldridge%
}{%
{\protect \APACyear {2022}}%
}]{%
wooldridge2022}
\APACinsertmetastar {%
wooldridge2022}%
\begin{APACrefauthors}%
Wooldridge, J\BPBI M.%
\end{APACrefauthors}%
\unskip\
\newblock
\APACrefYearMonthDay{2022}{}{}.
\newblock
{\BBOQ}\APACrefatitle {Simple approaches to nonlinear difference-in-differences
  with panel data} {Simple approaches to nonlinear difference-in-differences
  with panel data}.{\BBCQ}
\newblock
\APACjournalVolNumPages{Available at SSRN 4183726}{}{}{}.
\newblock
\begin{APACrefDOI} \doi{10.2139/ssrn.4183726} \end{APACrefDOI}
\PrintBackRefs{\CurrentBib}

\end{thebibliography}
	\vspace*{.5cm}
	
	\begin{literatur}
		\lititem \textbf{Data:}
	\end{literatur}
	
	\begin{literatur}
		\lititem Graf, T., Grie{\ss}emer, S., K\"ohler, M., Lehnert, C., Moczall, A., Oertel, M., Schmucker, A., Schneider, A., Seth, S., Thomsen, U., vom Berge, P. (2023). Weakly anonymous Version of the Sample of Integrated Labour Market Biographies (SIAB) - Version 7521 v1. \textit{Research Data Centre of the Federal Employment Agency (BA) at the Institute for Employment Research (IAB).} doi: \hyperlink{10.5164/IAB.SIAB7521.de.en.v1}{10.5164/IAB.SIAB7521.de.en.v1} \\
		\textit{The data access was provided via on-site use at the Research Data Centre (FDZ) of the German Federal Employment Agency (BA) at the Institute for Employment Research (IAB)  and subsequently remote data access.}
	\end{literatur}
	
	\begin{literatur}
		\lititem Regionaldatenbank (2024a). Niederlassungen nach Besch\"aftigtengr\"o{\ss}enklassen - Jahr - regionale Tiefe: Kreise und krfr. St\"adte (bis 2018). (52111-01-01-4)   
	\end{literatur}

	\begin{literatur}
		\lititem Regionaldatenbank (2024b). Zu- und Fortz\"uge (\"uber Gemeindegrenzen) nach Geschlecht und Altersgruppen - Jahressumme - regionale Tiefe: Kreise und krfr. St\"adte. (12711-01-03-4) 
	\end{literatur}
	
	
	\clearpage
	\begin{appendices}
		\setcounter{page}{1}
		\pagenumbering{roman}

		\section{Distribution of dependent variables}\label{app:apx_dist}
		
		\setcounter{figure}{0}
		\setcounter{table}{0}
		\renewcommand*{\thefigure}{\thesection\arabic{figure}}
		\renewcommand*{\thetable}{\thesection\arabic{table}}
		
		\begin{figure}[tbh]
			\caption{Distribution of inflows and outflows (2010 - 2019)}\label{fig:dist1_apx}
			\centering
			\includegraphics[width=0.84\textwidth]{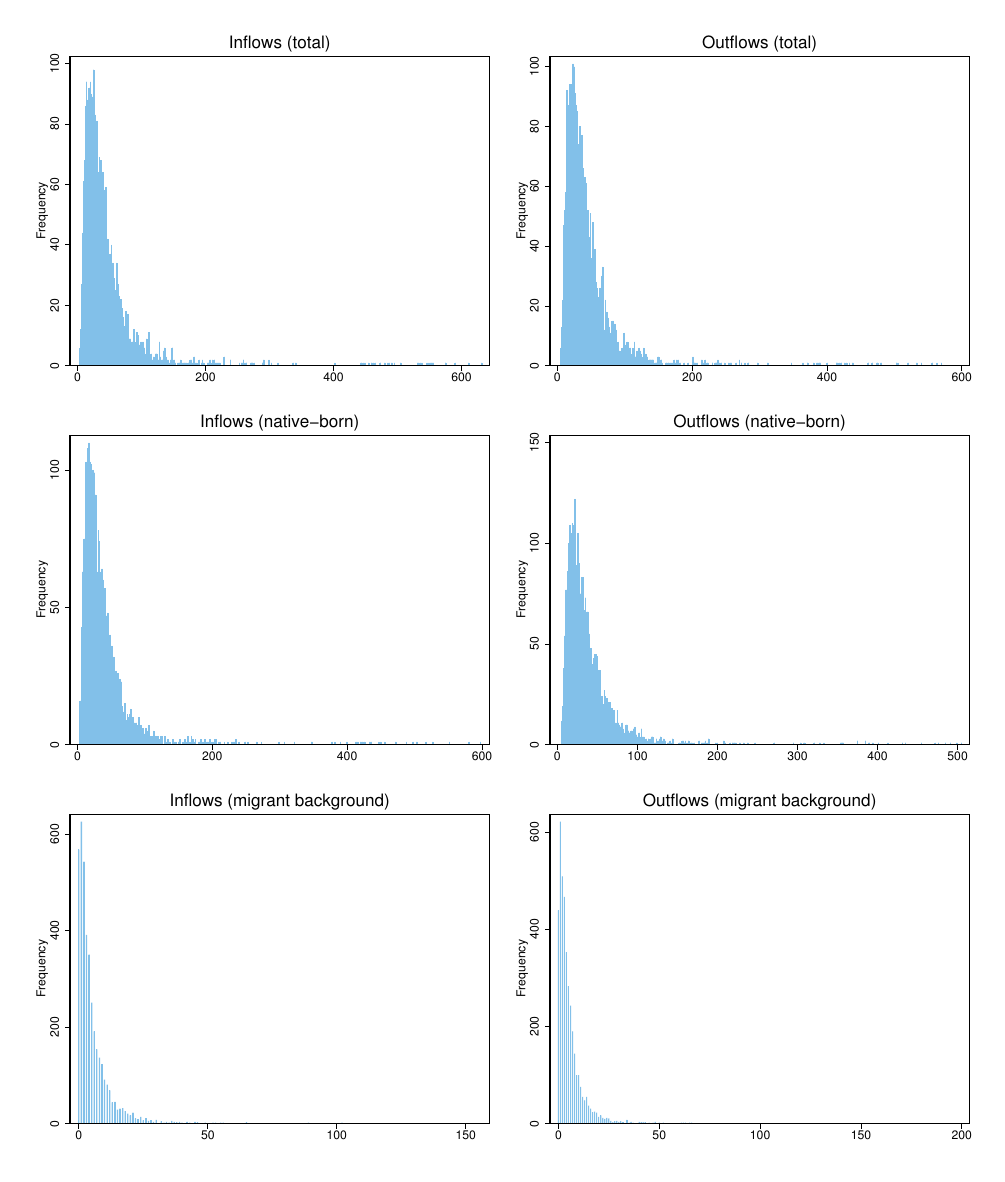}
			\begin{singlespace}
				\justifying \noindent \footnotesize \textit{Notes}:
				This figure shows the distribution of the dependent variables used in \fref{fig:baseline}. \textit{Source: SIAB, own calculations}
			\end{singlespace}
		\end{figure}
		
		\begin{figure}[tbh]
			\caption{Distribution of inflows and outflows (2010 - 2019)}\label{fig:dist2_apx}
			\centering
			\includegraphics[width=0.84\textwidth]{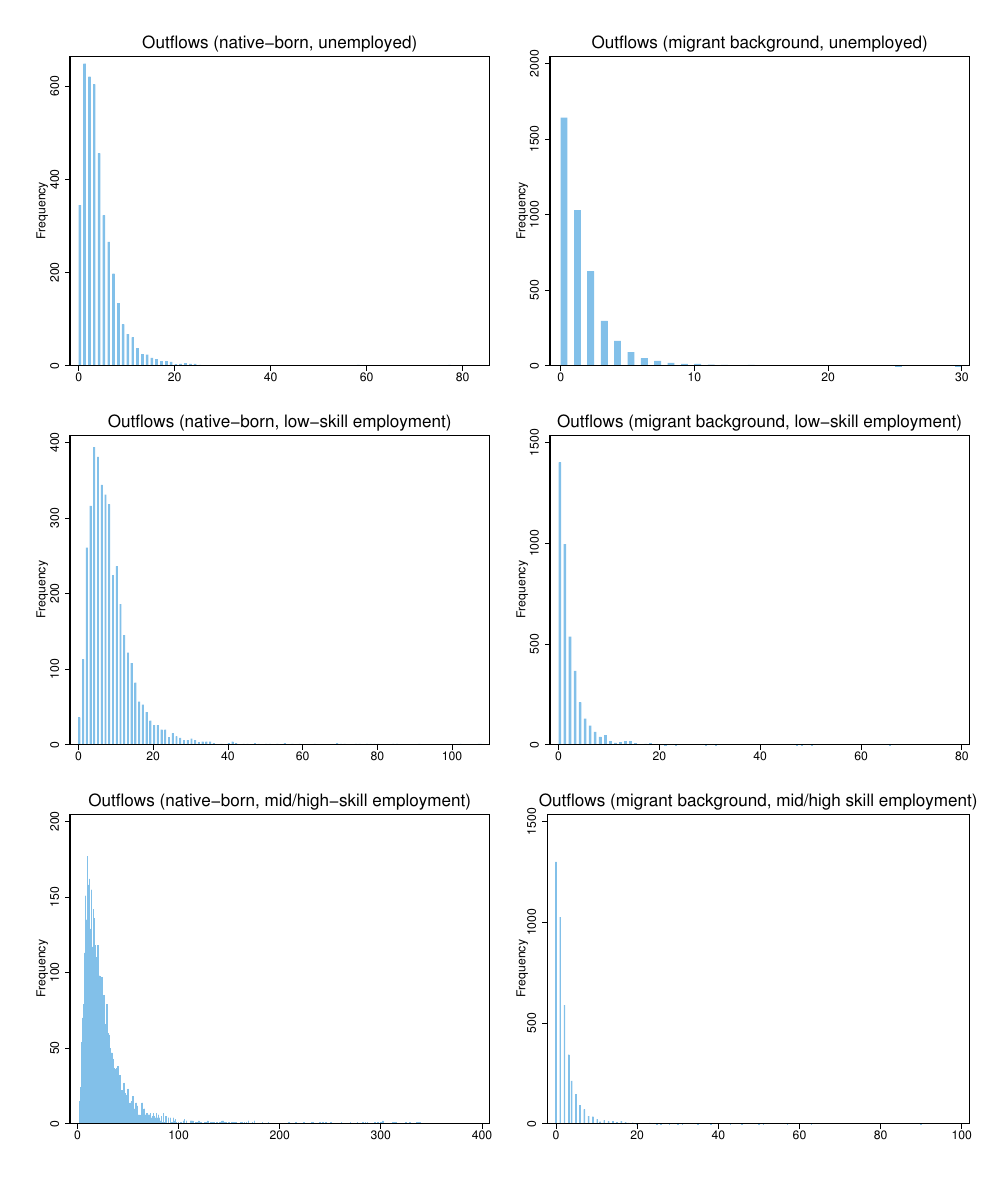}
			\begin{singlespace}
				\justifying \noindent \footnotesize \textit{Notes}:
				This figure shows the distribution of the dependent variables used in \fref{fig:het_employment}. \textit{Source: SIAB, own calculations}
			\end{singlespace}
		\end{figure}
		
		\begin{figure}[tbh]
			\caption{Distribution of inflows and outflows (2010 - 2019)}\label{fig:dist3_apx}
			\centering
			\includegraphics[width=0.84\textwidth]{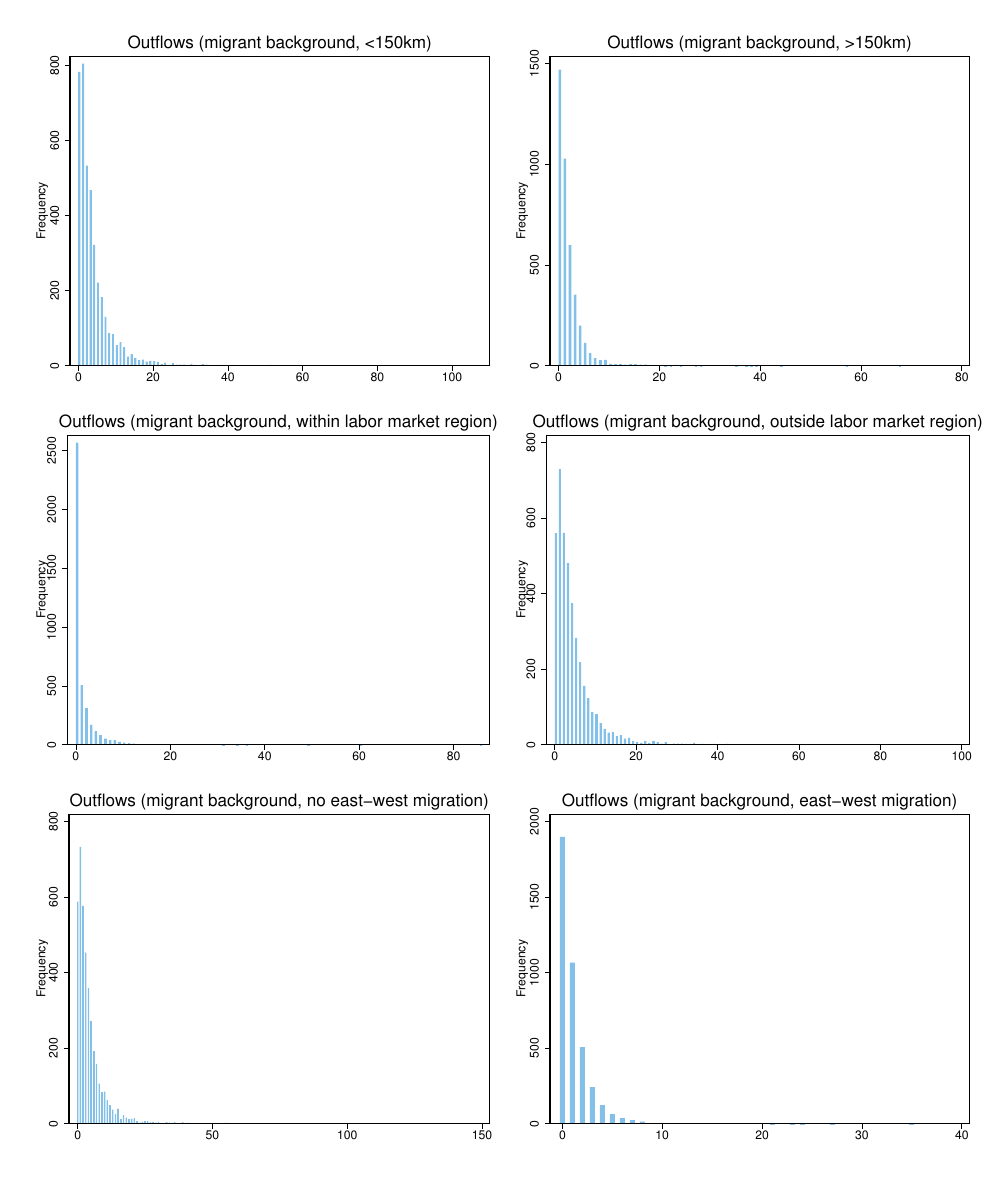}
			\begin{singlespace}
				\justifying \noindent \footnotesize \textit{Notes}:
				This figure shows the distribution of the dependent variables used in \fref{fig:het_dist}. \textit{Source: SIAB, own calculations}
			\end{singlespace}
		\end{figure}
		
		\section{Further results}\label{app:apx_res}
		
		\setcounter{figure}{0}
		\setcounter{table}{0}
		\renewcommand*{\thefigure}{\thesection\arabic{figure}}
		\renewcommand*{\thetable}{\thesection\arabic{table}}
		
\begin{figure}[ptbh]
			\caption{Treatment effects on outflows by regional variation}\label{fig:het_dist_mig0_apx}
			\centering
			\includegraphics[width=.9\textwidth]{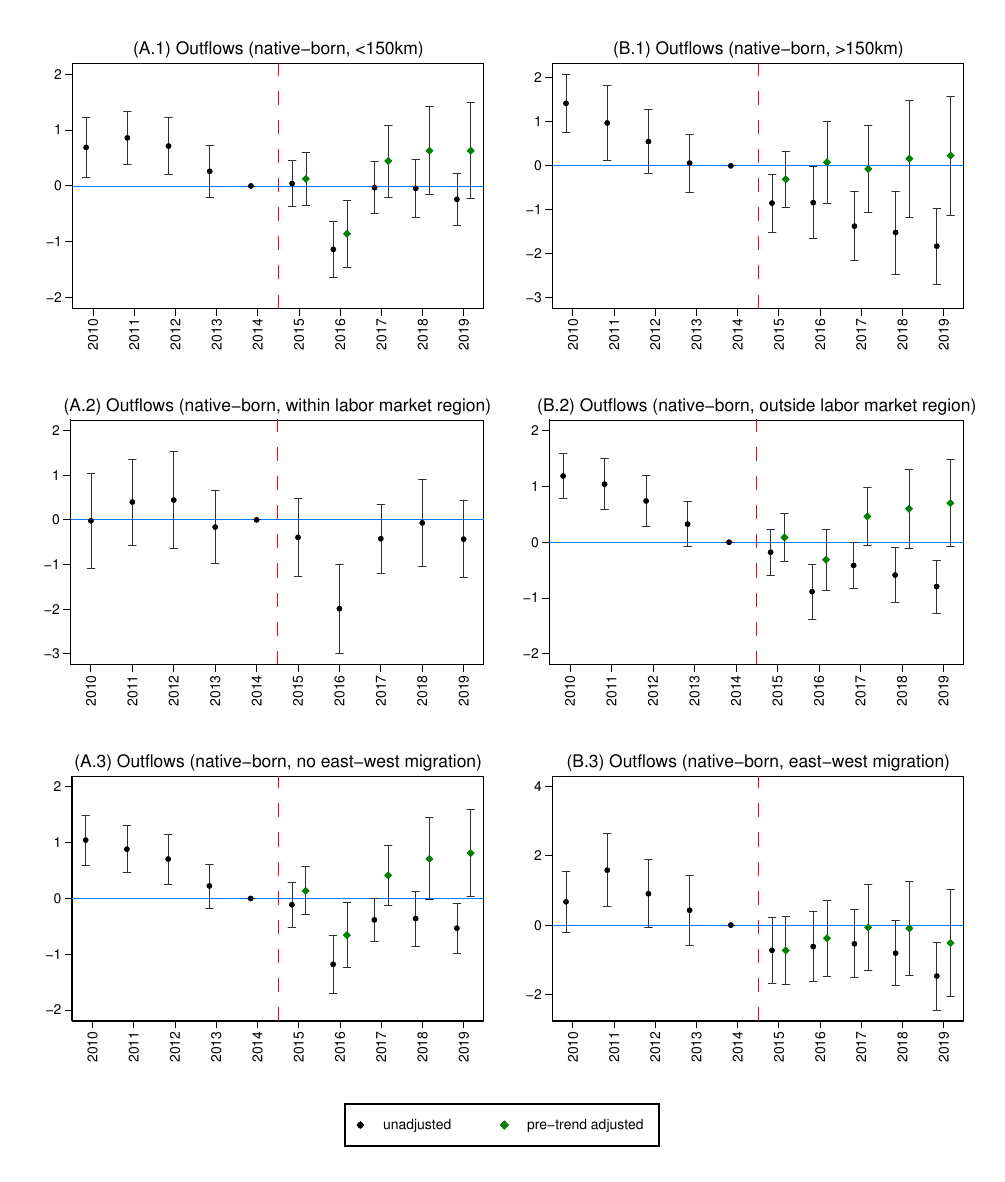} \\
			\begin{singlespace}
				\justifying \noindent \footnotesize \textit{Notes}: Cluster-robust standard errors are estimated and 95\% confidence intervals are displayed. $N=4,000$ over 400 districts. For specification (A.2) 170 districts are omitted due to no within-variation. Coefficients shown are $\hat{\gamma}_{t}$ from estimating a fixed effects Poisson model with the conditional mean specified as in \fref{eq:DIDposFE}. The pre-trend Wald test suggests a significant pre-trend for following specifications: (A.1), $p= 0.003$; (A.3), $p= 0.000$; (B.1), $p= 0.000$; (B.2), $p= 0.000$; (B.3), $p= 0.039$. For the respective specifications, trend-adjusted coefficients are estimated by specifying the conditional mean as in \fref{eq:DIDPOSFE_TA}. For all pre-trend corrections, $\hat{\pi}$ and the alternative estimate for the coefficient on $Bite_{s} \cdot Trend_{t}$ are very similar -- as described in \fref{sec:ident1} to check whether \fref{eq:DIDPOSFE_TA} sufficiently corrects for a linear pre-trend. The magnitude of the average treatment effects should be interpreted by multiplying $\hat{\gamma}_{t}$ with the average treatment intensity of about 10\%. \textit{Source: SIAB, own calculations}    
			\end{singlespace}
		\end{figure}

		\Fref{app:apx_res} provides estimates for additional subsamples in \fref{fig:het_dist_mig0_apx} according to \fref{eq:DIDposFE} and \fref{eq:DIDPOSFE_TA}.
		
		\clearpage
		
		\section{Further robustness checks}\label{app:apx_robust}
		
		\setcounter{figure}{0}
		\setcounter{table}{0}
		\renewcommand*{\thefigure}{\thesection\arabic{figure}}
		\renewcommand*{\thetable}{\thesection\arabic{table}}
		
		\Fref{app:apx_robust} provides results in \fref{tab:robust_margin_ls} to \fref{tab:robust_specs_unemp} on robustness checks as presented in \fref{tab:robust_margin} and \fref{tab:robust_specs} in \fref{sec:robust1} for further relevant sub-samples. 
		
		\clearpage
		
		\begin{table}[p]
			\centering
			\caption{Robustness check: Estimates on outflows including marginal employment for the sample of low-skilled workers}\label{tab:robust_margin_ls}
			\begin{tabularx}{\textwidth}{l*{4}{D{.}{.}{3.2}}}
				\toprule
				& \multicolumn{2}{c}{migrant background} & \multicolumn{2}{c}{native-born} \\
				\cmidrule(lr){2-3} \cmidrule(lr){4-5}
				& \multicolumn{1}{c}{w/o marg. empl.} & \multicolumn{1}{c}{w/ marg. empl.} & \multicolumn{1}{c}{w/o marg. empl.} & \multicolumn{1}{c}{w/ marg. empl.} \\
				\midrule
				$\hat{\gamma}_{2010}$ & 3.13^{***} & 2.77^{***} & 0.28 & 0.73^{*} \\
				$\hat{\gamma}_{2011}$ & 1.76 & 1.36 & 0.06 & 0.39 \\
				$\hat{\gamma}_{2012}$ & 3.07^{*} & 2.61^{*} & 0.23 & 0.39 \\
				$\hat{\gamma}_{2013}$ & 0.76 & 0.55 & -0.06 & 0.02 \\
				$\hat{\gamma}_{2014}$ & & & &  \\
				$\hat{\gamma}_{2015}$ & 1.82 & 0.48 & -0.55 & -0.38 \\
				$\hat{\gamma}_{2016}$ & 2.02^{*} & 1.96^{*} & -1.61^{***} & -1.32^{***} \\
				$\hat{\gamma}_{2017}$ & 2.65^{**} & 2.17^{**} & -0.72^{*} & -0.42 \\
				$\hat{\gamma}_{2018}$ & 2.79^{***} & 2.37^{**} & -0.4 & -0.07 \\
				$\hat{\gamma}_{2019}$ & 4.6^{***} & 3.87^{***} & -0.81^{*} & -0.47 \\
				\bottomrule
			\end{tabularx}
			\vspace{1mm}
			\begin{singlespace}
				\justifying \noindent \footnotesize \textit{Notes}: Estimates for $\gamma_t$ are provided for the low-skilled employed sub-samples without marginal employment and as a sensitivity analysis with marginal employment. $\hat{\gamma}_t$ are estimated with a conditional fixed effects Poisson model with the conditional mean specified as in \fref{eq:DIDposFE}. $\hat{\gamma}_{2014}$ is the base category. Cluster robust standard errors are estimated. $^{*} p < 0.10$, $^{**} p < 0.05$, $^{***} p < 0.01$. \textit{Source: SIAB, own calculations}
			\end{singlespace}
		\end{table}
		
		\clearpage
		
		\begin{table}[p]
			\centering
			\caption{Robustness check: Estimates on outflows including marginal employment for the sample of unemployed}\label{tab:robust_margin_unemp}
			\begin{tabularx}{\textwidth}{l*{4}{D{.}{.}{3.2}}}
				\toprule
				& \multicolumn{2}{c}{migrant background} & \multicolumn{2}{c}{native-born} \\
				\cmidrule(lr){2-3} \cmidrule(lr){4-5}
				& \multicolumn{1}{c}{w/o marg. empl.} & \multicolumn{1}{c}{w/ marg. empl.} & \multicolumn{1}{c}{w/o marg. empl.} & \multicolumn{1}{c}{w/ marg. empl.} \\
				\midrule
				$\hat{\gamma}_{2010}$ & -0.34 & -0.68 & 0.74 & 0.84 \\
				$\hat{\gamma}_{2011}$ & -0.85 & -1.47 & 1.46^{**} & 1.44^{***} \\
				$\hat{\gamma}_{2012}$ & 2.54 & 1.26 & 0.77 & 0.78 \\
				$\hat{\gamma}_{2013}$ & -0.70 & -0.98 & 0.76 & 0.76 \\
				$\hat{\gamma}_{2014}$ & & & &  \\
				$\hat{\gamma}_{2015}$ & 1.08 & -0.21 & -0.26 & -0.41 \\
				$\hat{\gamma}_{2016}$ & 2.11 & 1.17 & -0.02 & 0.06 \\
				$\hat{\gamma}_{2017}$ & 3.15^{***} & 2.10^{*} & -0.24 & -0.18 \\
				$\hat{\gamma}_{2018}$ & 3.01^{**} & 2.06^{*} & -0.78 & -0.74 \\
				$\hat{\gamma}_{2019}$ & 3.26^{***} & 2.51^{**} & -0.21 & -0.13 \\
				\bottomrule
			\end{tabularx}
			\vspace{1mm}
			\begin{singlespace}
				\justifying \noindent \footnotesize \textit{Notes}: Estimates for $\gamma_t$ are provided for the unemployed sub-samples without marginal employment and as a sensitivity analysis with marginal employment. $\hat{\gamma}_t$ are estimated with a conditional fixed effects Poisson model with the conditional mean specified as in \fref{eq:DIDposFE}. $\hat{\gamma}_{2014}$ is the base category. Cluster robust standard errors are estimated. $^{*} p < 0.10$, $^{**} p < 0.05$, $^{***} p < 0.01$. \textit{Source: SIAB, own calculations}
			\end{singlespace}
		\end{table}
		
		\clearpage
		
		\begin{table}[p]
			\centering
			\caption{Robustness check: Estimates on outflows varying between count data models for the sample of low-skilled workers} \label{tab:robust_specs_ls}%
			\begin{tabularx}{\textwidth}{l*{6}{D{.}{.}{3.2}}}
				\toprule
				& \multicolumn{3}{c}{migrant background} & \multicolumn{3}{c}{native-born} \\
				\cmidrule(lr){2-4} \cmidrule(lr){5-7}
				& \multicolumn{1}{c}{FE Poisson} & \multicolumn{1}{c}{pooled Poisson} & \multicolumn{1}{c}{log-linear} & \multicolumn{1}{c}{FE Poisson} & \multicolumn{1}{c}{pooled Poisson} & \multicolumn{1}{c}{log-linear} \\
				\midrule
				$\hat{\gamma}_{2010}$ & 3.13^{***} & 2.87^{***} & 16.08 & 0.28 & 0.27 & -0.34 \\
				$\hat{\gamma}_{2011}$ & 1.76 & 1.61 & 16.57 & 0.06 & 0.06 & -0.08 \\
				$\hat{\gamma}_{2012}$ & 3.07^{*} & 2.82^{*} & 25.77^{**} & 0.23 & 0.22 & 1.7 \\
				$\hat{\gamma}_{2013}$ & 0.76 & 0.69 & 13.95 & -0.06 & -0.05 & -0.6 \\
				$\hat{\gamma}_{2014}$ &  &  &  &  &  &  \\
				$\hat{\gamma}_{2015}$ & 1.82 & 1.66 & 9.54 & -0.55 & -0.52 & -1.31^{*} \\
				$\hat{\gamma}_{2016}$ & 2.02^{*} & 1.84^{*} & 22.66^{**} & -1.61^{***} & -1.51^{***} & -3.84 \\
				$\hat{\gamma}_{2017}$ & 2.65^{**} & 2.42^{**} & 11.1 & -0.72^{*} & -0.68^{*} & -5.8 \\
				$\hat{\gamma}_{2018}$ & 2.79^{***} & 2.55^{***} & 17.18^{*} & -0.4 & -0.38 & -1.51 \\
				$\hat{\gamma}_{2019}$ & 4.6^{***} & 4.24^{***} & 33.51^{***} & -0.81^{*} & -0.76^{*} & -3.3 \\
				\bottomrule
			\end{tabularx}
			\vspace{1mm}
			\begin{singlespace}
				\justifying \noindent \footnotesize \textit{Notes}: Estimates for $\gamma_t$ are provided for the low-skilled employed sub-samples. $\hat{\gamma}_t$ are estimated in column (1) and (4) with a conditional fixed effects Poisson model with the conditional mean specified as in \fref{eq:DIDposFE}. In column (2) and (5) $\hat{\gamma}_{t}$ are estimated with a pooled Poisson model with the conditional mean specified as in \fref{eq:DIDpos}. In column (3) and (6) $hat{\gamma}_t$ is estimated with a log-linear model as in \fref{eq:loglin}. $\hat{\gamma}_{2014}$ is the base category. Cluster robust standard errors are estimated. $^{*} p < 0.10$, $^{**} p < 0.05$, $^{***} p < 0.01$.  \textit{Source: SIAB, own calculations}
			\end{singlespace}
		\end{table}
		
		\clearpage
		
		\begin{table}[p]
			\centering
			\caption{Robustness check: Estimates on outflows varying between count data models for the sample of unemployed} \label{tab:robust_specs_unemp}%
			\begin{tabularx}{\textwidth}{l*{6}{D{.}{.}{3.2}}}
				\toprule
				& \multicolumn{3}{c}{migrant background} & \multicolumn{3}{c}{native-born} \\
				\cmidrule(lr){2-4} \cmidrule(lr){5-7}
				& \multicolumn{1}{c}{FE Poisson} & \multicolumn{1}{c}{pooled Poisson} & \multicolumn{1}{c}{log-linear} & \multicolumn{1}{c}{FE Poisson} & \multicolumn{1}{c}{pooled Poisson} & \multicolumn{1}{c}{log-linear} \\
				\midrule
				$\hat{\gamma}_{2010}$ & -0.34 & -0.29 & 10.33 & 0.74 & 0.64 & -2.52 \\
				$\hat{\gamma}_{2011}$ & -0.85 & -0.72 & 15.19 & 1.46^{**} & 1.28^{**} & 6.35 \\
				$\hat{\gamma}_{2012}$ & 2.54 & 2.20 & 20.49^{*} & 0.77 & 0.68 & 9.00 \\
				$\hat{\gamma}_{2013}$ & -0.70 & -0.59 & 14.03 & 0.76 & 0.66 & 3.88 \\
				$\hat{\gamma}_{2014}$ &  &  &  &  &  &  \\
				$\hat{\gamma}_{2015}$ & 1.08 & 0.93 & 14.96 & -0.26 & -0.23 & -11.86 \\
				$\hat{\gamma}_{2016}$ & 2.11 & 1.82 & 30.74^{***} & -0.02 & -0.02 & -7.27 \\
				$\hat{\gamma}_{2017}$ & 3.15^{***} & 2.73^{**} & 28.64^{**} & -0.24 & -0.21 & -8.90 \\
				$\hat{\gamma}_{2018}$ & 3.01^{**} & 2.61^{**} & 17.94 & -0.78 & -0.67 & -1.87 \\
				$\hat{\gamma}_{2019}$ & 3.26^{***} & 2.84^{***} & 37.38^{***} & -0.21 & -0.18 & -7.13 \\
				\bottomrule
			\end{tabularx}
			\vspace{1mm}
			\begin{singlespace}
				\justifying \noindent \footnotesize \textit{Notes}: Estimates for $\gamma_t$ are provided for the unemployed sub-samples. $\hat{\gamma}_t$ are estimated in column (1) and (4) with a conditional fixed effects Poisson model with the conditional mean specified as in \fref{eq:DIDposFE}. In column (2) and (5) $\hat{\gamma}_{t}$ are estimated with a pooled Poisson model with the conditional mean specified as in \fref{eq:DIDpos}. In column (3) and (6) $\hat{\gamma}_t$ is estimated with a log-linear model as in \fref{eq:loglin}. $\hat{\gamma}_{2014}$ is the base category. Cluster robust standard errors are estimated. $^{*} p < 0.10$, $^{**} p < 0.05$, $^{***} p < 0.01$.  \textit{Source: SIAB, own calculations}
			\end{singlespace}
		\end{table}
		
	\end{appendices}
	
\end{document}